\documentclass[lettersize,journal]{IEEEtran}
\usepackage{amsmath,amsfonts}
\usepackage{array}
\usepackage[caption=false,font=normalsize,labelfont=sf,textfont=sf]{subfig}
\usepackage{textcomp}
\usepackage{stfloats}
\usepackage{url}
\usepackage{verbatim}
\usepackage{graphicx}
\usepackage{siunitx}

\usepackage{bbm}
\usepackage{bm}
\usepackage{amsmath,amsfonts}
\usepackage{amsthm}

\newtheorem{definition}{Definition}

\usepackage{slashed}
\usepackage{booktabs}
\usepackage{subfig}
\usepackage{algorithm}
\usepackage[noend]{algpseudocode}

\usepackage{color}
\definecolor{XFcolor}{RGB}{0,0,0}
\definecolor{XFcolorb}{RGB}{25,202,173}
\usepackage[braket]{qcircuit}

\usepackage{cite}
\begin{document}

\title{Gate-Based and Annealing-Based Quantum Algorithms for the Maximum K-Plex Problem}


\author{\IEEEauthorblockN{Xiaofan Li\IEEEauthorrefmark{2},
Gao Cong\IEEEauthorrefmark{2},
 and
Rui Zhou\IEEEauthorrefmark{3}}

\IEEEauthorblockA{
\IEEEauthorrefmark{2}Nanyang Technological University, Singapore. Email: \{xiaofan.li,gaocong\}@ntu.edu.sg\\
\IEEEauthorrefmark{3}
Swinburne University of Technology, Australia.  
Email: rzhou@swin.edu.au
}
}

\markboth{Journal of \LaTeX\ Class Files,~Vol.~14, No.~8, August~2021}%
{Shell \MakeLowercase{\textit{et al.}}: A Sample Article Using IEEEtran.cls for IEEE Journals}


\maketitle

\begin{abstract}
The \( k \)-plex model, which allows each vertex to miss connections with up to \( k \) neighbors, serves as a relaxation of the clique. Its adaptability makes it more suitable 
for analyzing real-world graphs 
where noise and imperfect data are common and the ideal clique model is often impractical.
The problem of identifying the maximum \( k \)-plex (MKP, which is NP-hard) is gaining attention in fields such as social network analysis, community detection, terrorist network identification, and graph clustering. Recent works have focused on optimizing the time complexity of MKP algorithms. The state-of-the-art  has reduced the complexity from a trivial \( O^*(2^n) \) to \( O^*(c_k^n) \), with \( c_k > 1.94 \) for \( k \geq 3 \), where \( n \) denotes the vertex number.
This paper investigates the MKP using two quantum models: gate-based model and  annealing-based model. Two gate-based   algorithms, qTKP and qMKP, are proposed to achieve \( O^*(1.42^n) \) time complexity. qTKP integrates quantum search with graph encoding, degree counting, degree comparison, and size determination to find  a \( k \)-plex of a given size; qMKP uses binary search to progressively  identify the maximum solution. 
Furthermore, by reformulating MKP as a quadratic unconstrained binary optimization problem, we propose qaMKP, the first   annealing-based approximation algorithm, which utilizes qubit resources more efficiently than gate-based algorithms.
To validate the practical performance, proof-of-principle experiments were conducted using the latest IBM gate-based quantum simulator and D-Wave adiabatic quantum computer. 
This work holds potential  to be applied to a wide range of  clique relaxations, e.g.,  \( n \)-clan and \( n \)-club. 
\end{abstract}


\section{Introduction}

\begin{figure}[t]

	\centering
	{\includegraphics[width=5.5cm]{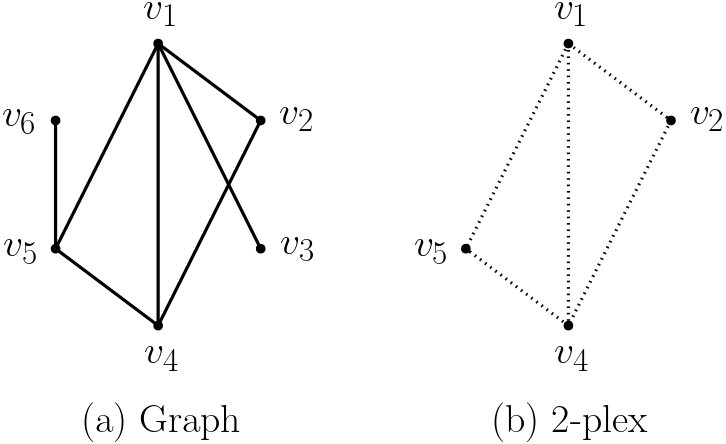}}

	\vspace{-8pt}
	\captionsetup{justification=centering}
	\caption{A graph and a $2$-plex}\label{graph}
	\vspace{-20pt}
\end{figure}

\textbf{Problem.}
Given an undirected, unweighted graph \( G = (V, E) \),  a \( k \)-plex in this graph is a vertex subset \( P \subseteq V \) where each vertex in \( P \) connects to at least \( |P| - k \) neighbors in \( P \). 
Figure~\ref{graph} shows an example.  
The maximum \( k \)-plex problem (MKP) is to find the largest \( k \)-plex  in $G$ for a given \( k \).

\textbf{Significance.} 
Identifying large cohesive subgraphs is critical in social network analysis and is relevant across domains such as social media, collaboration networks, communication networks, and web graphs. The \(k\)-plex structure is a widely used cohesive subgraph model. The search for the largest \(k\)-plex is widely used in social network analysis~\cite{xiao2017fast,conte2018d2k,gao2018exact}, community detection~\cite{conte2018d2k,zhou2020enumerating,zhu2020community}, terrorist network detection~\cite{krebs2002mapping}, and graph clustering~\cite{du2007community,newman2001structure}. The \(k\)-plex model is a relaxed version of a clique~\cite{pattillo2013clique}, where a clique seeks complete subgraphs and represents an ideal in cohesive subgraph detection. In real-world networks, data often contain noise and errors, making large perfect cliques rare~\cite{pattillo2013clique}. The \(k\)-plex concept allows each vertex in a subgraph to miss up to \(k\) neighbors, making this model more suitable for real-world network structures than the strict clique model.

\textbf{Uncharted opportunity.}
The maximum \( k \)-plex problem is NP-hard~\cite{lewis1980node}, and under the assumption \(P \neq NP\), no polynomial-time algorithm can approximate the solution within a factor better than \( O(n^\delta) \) for any \( \delta > 0 \), where \( n = |V| \)~\cite{lund1993approximation}. Due to its complexity, branch-and-bound methods, despite their exponential worst-case time, are often used to develop exact algorithms. These approaches adapt co-\( k \)-plex coloring for upper bounds~\cite{mcclosky2012combinatorial}, employ dynamic vertex selection for graph reduction~\cite{gao2018exact}, design coloring-based~\cite{zhou2021improving} and partition-based~\cite{jiang2021new} upper bounds, and use matrix-based methods with first-order and second-order pruning~\cite{chang2022efficient}. However, these algorithms still exhibit a worst-case time complexity of \( O^*(2^n) \), where \( O^*(\cdot) \) hides polynomial factors.
Researchers continue to seek breakthroughs in complexity optimization. For \( k = 1 \), MKP can be solved in \( O^*(1.1996^n) \) as of 2013~\cite{xiao2013exact}; for \( k = 2 \), in \( O^*(1.3656^n) \) as of 2016~\cite{xiao2017exact}; and for \( k \geq 3 \), in \( O(c_k^n n^{O(1)}) \) as of 2017~\cite{xiao2017fast}, where \( c_k \) is an increasing sequence with \( c_3 = 1.9476, c_4 = 1.9786, c_5 = 1.9910, \ldots \). For \( k > 2 \), although these algorithms surpass the trivial \( O^*(2^n) \) bound, the exponential base \( c_k \) remains close to 2. The challenge in further optimizing MKP time complexity may stem from a lack of suitable mathematical structure in its solution space, which might contain \( \Omega^*(c_k^n) \) elements. In such an unstructured space, traditional search methods are unlikely to reduce complexity below \( O^*(c_k^n) \). Therefore, new algorithms based on quantum search principles are needed for further optimization. 

Quantum computing (QC) is considered to excel at   optimization problems that require identifying elements with specific properties within exponentially large search spaces, a common challenge in database and graph database systems. Recent research suggests that quantum computing achieves quadratic speed-ups in time complexity over classical algorithms for many NP-hard problems~\cite{gaitan2014graph,su2016quantum,srinivasan2018efficient}. 
However, QC has seen limited adoption in graph database research, primarily due to the abstract nature of designing quantum algorithms~\cite{nielsen_chuang_2010} and the challenge of representing complex graph problems on state-of-the-art architectures, such as gate-based and annealing-based models. 
For the MKP, the gate-based model requires encoding the graph topology into quantum memory and designing quantum circuits to verify if a subgraph qualifies as a \(k\)-plex, while the annealing-based model requires transforming the MKP into a quadratic unconstrained binary optimization form solvable by an adiabatic quantum computer and improving the utilization of qubit resources. In this work, we fill these gaps.

\textbf{Our approach.}
In this study, we explore how the quantum computing can be harnessed to solve graph database problems, specifically, the maximum $k$-plex problem. 
We propose two types of quantum algorithms based on state-of-the-art architectures: gate-based  qTKP and qMKP with quantum circuit design, and  annealing-based   qaMKP, proving the following fact: 
{for any given positive integer \( k \), MKP can be solved in \( \mathbf{O^*(2^{n/2})} \).}
The key to the gate-based algorithms qTKP/qMKP is translating MKP into quantum logic using quantum circuits and identifying the solution in the search space via a black box, which is termed as the {\it oracle}. 
We design a graph encoding circuit and a three-part quantum oracle for solution identification, with functions of degree count, degree comparison, and size determination. 
qTKP finds a \( k \)-plex of at least size \( T \) in a given graph, 
based on which  
 qMKP employs a binary search to locate the maximum \( k \)-plex, progressively yielding a feasible solution of at least half the optimal size within the first \( O(1/\log n) \) of the running time. Both algorithms have a time complexity of \( O^*(2^{n/2}) \).
The key to the annealing-based algorithm qaMKP is transforming MKP into a quadratic unconstrained binary optimization (QUBO) form~\cite{10.14778/2947618.2947621,schonberger2022applicability}. This is to find an unconstrained quadratic energy objective function with only binary variables such that the variables at the minimum energy state correspond to an MKP solution. We convert the inequality constraints on vertex degree into equality constraints by introducing \( O(n\log n) \) slack variables, then transform these into quadratic penalties added to the energy function, imposing penalties on solutions that violate constraints to filter out invalid ones. Compared with qTKP and qMKP, qaMKP achieves higher qubit resource utilization.

We highlight our main contributions below.
\begin{itemize}
\item Our work represents the pioneering effort to examine the Maximum $k$-Plex Problem  through a quantum perspective, providing new mathematical reformulations that consider specific properties and restrictions. The proposed approached hold  potential for application to various clique relaxations, including \(n\)-clan and \(n\)-club.
\item We proposed  two gate-based  algorithms   qTKP and qMKP, which   identify a $k$-plex with a given size and the maximum size, respectively. Both algorithms achieve a time complexity of \(O^*(2^{n/2})\), offering a near-quadratic speedup for \(k \geq 3\) compared to the state-of-the-art.
\item We proposed the first annealing-based  algorithm  qaMKP by transforming the problem into a QUBO form with only \(O(n\log n)\) slack variables. Compared with qTKP and qMKP, qaMKP exhibits higher qubit resource utilization.
\item We conducted proof-of-principle experiments using the latest gate-based quantum simulators and adiabatic quantum computers to validate the performance of our proposed algorithms.
\end{itemize}

\textbf{Roadmap.}
The remainder of the paper is organized as follows.
Section~\ref{sec:2} covers preliminaries.
Section~\ref{sec:3} introduces our gate-based algorithms qTKP and qMKP. 
Section~\ref{sec:ann} gives our annealing-based algorithm  qaMKP. 
Section~\ref{sec:4} conducts experimental studies. 
Section~\ref{sec:5} reviews related works. 
Conclusion and limitation are discussed in  Section~\ref{sec:6}.

\section{Preliminaries}\label{sec:2}
This section covers the preliminaries of 
$k$-plex and an overview of gates-based  quantum circuits. We also review Grover's algorithm,  
a fundamental quantum search algorithm underpinning our gate-based methods. 

\subsection{Maximum $k$-Plex Problem (MKP)}

Consider an unweighted and undirected graph $G = (V, E)$, where $V$ is the set of vertices and $E$ is the set of edges. 
The degree of a vertex $u$ in subgraph $S$, denoted as $d_S(u)$, is quantified by the number of its neighborhoods in $S$. 

\begin{definition}[\textbf{$k$-Plex}]\label{kplex}
A vertex set $P\subseteq V$ is defined as a $k$-plex if each vertex $v\in P$ has a degree $d_P(v)\geq |P|-k$. Here, $k$ is an integer that satisfies $k\in [1,|P|-1]$.
\end{definition}
 
The value of \( k \) indicates the deviation from a complete clique, with a clique equating to a \( 1 \)-plex. The size of a \( k \)-plex \( P \) is defined by the number of its vertices, denoted as \( |P| \). Our research focuses on the following problem:


%

\begin{definition}[\textbf{Maximum \(k\)-Plex Problem (MKP)}]
	Given an unweighted and undirected graph \( G = (V, E) \) with an integer \( k \in  [1, |V|-1] \), the Maximum \( k \)-Plex Problem aims to find a \( k \)-plex  that has the greatest number of vertices. 
\end{definition}


\subsection{Quantum Computing}
In quantum computing, a quantum state is represented as a vector, with its time evolution governed by operations similar to vector rotations. In our context, we focus on the basic unit of quantum information, the qubit, which is defined as:

\begin{definition}[\textbf{Qubit}]
	A qubit is a unit-norm vector within a two-dimensional complex  space, represented as:
	\begin{equation}
		\ket{q} = \alpha \ket{0} + \beta \ket{1}
	\end{equation}
	In this expression, the symbol $\ket{\cdot}$ signifies a vector state. The vectors $\ket{0}$ and $\ket{1}$ form an orthonormal basis for this space. The complex numbers $\alpha$ and $\beta$ are termed amplitudes, which satisfies the normalization condition $|\alpha|^2 + |\beta|^2 = 1$.
\end{definition}


A qubit, unlike a classic bit that is either 0 or 1, exists in a superposition state \(\alpha \ket{0} + \beta \ket{1}\), representing a continuous blend of both states. This can be visualized as a resultant vector distinct from both of the base vectors. When measured, a qubit \(\ket{q}\) collapses to one of the base states: \(\ket{0}\) with probability \(|\alpha|^2\) or \(\ket{1}\) with probability \(|\beta|^2\), known as quantum measurement collapse.
When analyzing a system with \( n \) qubits, we use a tensor product to describe its state, e.g., the state of a two-qubit system is represented as:
%
\begin{equation*}
	\begin{aligned}
		\ket{q_{comp}} &= \ket{q_1} \ket{q_2}
		= (\alpha_1 \ket{0} + \beta_1 \ket{1})(\alpha_2 \ket{0} + \beta_2 \ket{1}) \\
		&= \alpha_1\alpha_2\ket{00} + \alpha_1\beta_2\ket{01}
		+ \alpha_2\beta_1\ket{10} + \beta_1\beta_2\ket{11}
	\end{aligned}
\end{equation*} 
Here, we denote the tensor product of two qubit states \(\ket{i}\) and \(\ket{j}\) as \(\ket{ij}\). Quantum state evolution, characterized by rotations in vector space, is mathematically represented through matrix multiplication:
\begin{equation}
	\ket{q_{initial}}\xrightarrow[]{\mbox{over time}}\ket{q_{final}} = U\ket{q_{initial}}
\end{equation}
The matrix \( U \) is unitary, satisfying \( U^{\dag}U = I \), where \( U^{\dag} \) is the conjugate transpose of \( U \), and \( I \) represents the identity matrix. For instance, the matrix \( X \) flips the state of a qubit from \(\ket{0}\) to \(\ket{1}\) and vice versa:
\begin{equation}\label{eq:X}
	X\ket{q} = \alpha X\ket{0} + \beta X\ket{1} = \alpha \ket{1} + \beta \ket{0}
\end{equation}
This matrix $X$ is known as the quantum NOT gate. 
%
%
Another matrix we frequently use in this work  is the Hadamard matrix, denoted as $H$. This matrix transforms the state \(\ket{0}\) into a balanced superposition \((\ket{0} + \ket{1})/\sqrt{2}\), and similarly converts \(\ket{1}\) into \((\ket{0} - \ket{1})/\sqrt{2}\).

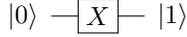
\begin{figure}[t]
\vspace{-12pt}
\[ \Qcircuit @C=1em @R=.7em {
      \lstick{\ket{0}}& \gate{X} & \rstick{\ket{1}}\qw
}\]
\vspace{-13pt}
\caption{A toy quantum circuit}
\label{fig:xgate}
\end{figure}

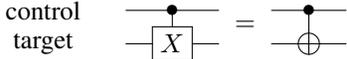
\begin{figure}[t]
\vspace{-8pt}
\[ \Qcircuit @C=1em @R=.5em {
    \hspace{-2.2cm} \mbox{control} & \ctrl{1} & \qw & && \ctrl{1} & \qw \\
     \hspace{-2.2cm}\mbox{target} & \gate{X} & \qw &\ustick{=} && \targ & \qw 
}\]
\vspace{-9pt}
\caption{Two representations of a CNOT gate}
\label{fig:cnot}
\vspace{-10pt}
\end{figure}



\subsection{Computation Model: Quantum Circuit}

A quantum circuit diagram displays the evolution of quantum states over time, consisting of qubits (quantum wires) and unitary gates that modify them. For example, Figure~\ref{fig:xgate} shows a basic circuit where an \(X\) gate transitions a qubit from state \(\ket{0}\) to \(\ket{1}\); the horizontal progression indicates temporal order. The \(X\) gate functions like a classical NOT gate by flipping a qubit's state, but it can also act on a qubit in superposition, unlike the classical NOT which works only on definite 0 or 1 states (see Eq.~\ref{eq:X}).


\begin{figure}[t]
\vspace{-12pt}
\[ \Qcircuit @C=1em @R=.5em {
    \hspace{-1.8cm} \mbox{control} & \ctrlo{1} & \qw & &&\hspace{10pt}\mbox{control} &&&&\ctrl{1} & \qw \\
     \hspace{-1.8cm}\mbox{target} & \targ & \qw & &&  \hspace{10pt}\mbox{control} &&&&\ctrl{1} & \qw \\
     &&&&&\hspace{10pt}\mbox{target} &&&&\targ & \qw
}\]
\vspace{-8pt}
\caption{Other types of controlled-gate}
\label{control}
\end{figure}
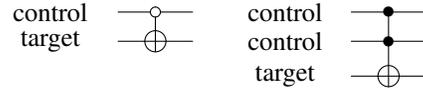

\begin{figure*}[t]
\vspace{-10pt}

	\centering
	{\includegraphics[width=18cm]{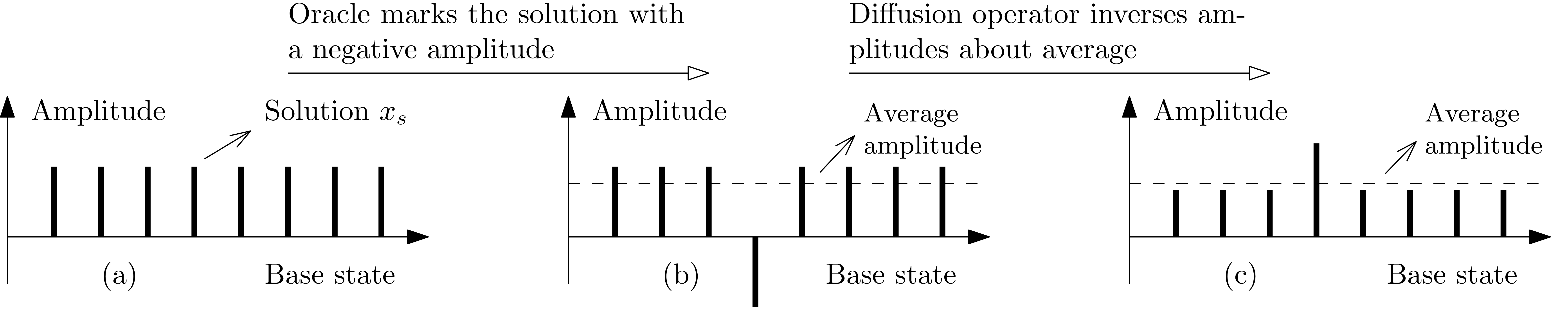}}

	\vspace{-8pt}
	\captionsetup{justification=centering}
	\caption{Illustration of the Grover's search with $n = 3$}\label{grover}
	\vspace{-5pt}
\end{figure*}

In quantum computing, the controlled-\(X\) (CNOT) gate is a fundamental logic gate (Fig.~\ref{fig:cnot}). It uses a control qubit (filled dot) and a target qubit, where the \(X\) operation applies to the target only if the control is in state \(\ket{1}\); otherwise, the target remains unchanged. For clarity, the \(X\) gate is depicted as a circle with a cross, and a hollow circle indicates inversion when the control is \(\ket{0}\). Controlled operations can extend to multiple control qubits, with the target acted upon only when all controls are in specific states. Such operations are known as C\(^k\)NOT gates, as shown in Figure~\ref{control}.

\subsection{Grover's Search}

Our gate-based algorithms follow the structure of the  Grover's search algorithm. 
It is initially designed for searching within an unstructured database, which is defined as:

\begin{definition}[\textbf{Unstructured Database Search}]
Consider the set \(\mathcal X = \{0, 1, \ldots, 2^{n}-1\}\) consisting of \(2^{n}\) integers. Let a function \(f: \mathcal X \rightarrow \{0,1\}\) have such that a unique element \(x_s \in \mathcal X\) satisfies \(f(x_s) = 1\), and for any other element \(x \in \mathcal X\) with \(x \neq x_s\), \(f(x) = 0\). The goal is to find \(x_s\).
\end{definition}

%

Each integer \(x \in \mathcal{X}\) can be represented as an \(n\)-bit binary string, i.e., as a tensor product of \(n\) qubit states. For example, the integer 3 is denoted \(\ket{0\ldots011}\) or simply \(\ket{3}\). The strategy operates on all \(2^n\) base states simultaneously in superposition. The algorithm iteratively amplifies the probability amplitude of the target state \(\ket{x_s}\), increasing it relative to non-solution states. The steps are detailed in Algorithm~\ref{algo:grover}.

\begin{algorithm}[ttttt]
\caption{Grover's Search Algorithm}
\begin{algorithmic}[1]
\Require Set of integers $\mathcal X = \{0, 1, \ldots, 2^{n}-1\}$ and function $f: \mathcal X\rightarrow \{0,1\}$ identifying a unique solution.
\Ensure Integer $x_s$ satisfying $f(x_s) = 1$. 

\State Prepare an equal superposition of all states $\frac{1}{\sqrt{2^n}}\sum_{i=0}^{2^n-1}\ket{i}$, each with equal probability.
\State Use a black box to flip the amplitude of solution state $\ket{x_s}$ from $+\frac{1}{\sqrt{2^n}}\ket{x_s}$ to $-\frac{1}{\sqrt{2^n}}\ket{x_s}$.
\State Apply the diffusion operator to amplify  the marked state by inverting amplitudes about their average.
\State Repeat steps 2 and 3 for $\lfloor\frac{\pi}{4}\sqrt{2^n}\rfloor$ iterations.
\State Measure the system to obtain the binary string  $x_s$.
\end{algorithmic}
\label{algo:grover}
\end{algorithm}

%

\textbf{Explanation:}

1. The equal superposition state is prepared by using $n$ $H$ gates to act on 
	$n$ initial state $\ket{0}$s: 
	\begin{equation*}
    \underbrace{(H\ket{0})(H\ket{0})\ldots(H\ket{0})}_{n \text{ times}} = \frac{1}{\sqrt{2^n}}\sum_{i=0}^{2^n-1}\ket{i}.
\end{equation*}		
Figure~\ref{grover}a visualizes this state with a bar graph. The horizontal axis represents different basis vectors, and the vertical axis shows each state's amplitude. For \(n = 3\), there are \(2^3 = 8\) basis states, each shown as a bar. In an equal superposition, all bars have the same height, indicating equal amplitudes.
2. The black box, or oracle, identifies the solution state \(\ket{x_s}\) (see Figure~\ref{grover}b). The diagram shows the solution state's amplitude inverted below the x-axis, indicating a phase inversion (amplitude multiplied by -1), while other states remain unchanged. The average amplitude, including the inverted solution state, becomes slightly lower than the non-solution states' amplitudes. This average is shown as a dashed line in the graph, just below the height of the non-solution bars.

3. Let \(\alpha\) be any amplitude and \(\overline{\alpha}\) the average amplitude. The diffusion operator inverts \(\alpha\) about the average, changing it to \(2\overline{\alpha} - \alpha\) (see Figure~\ref{grover}c). Compared to Figure~\ref{grover}a, after applying the diffusion operator, non-solution state amplitudes decrease while the solution state's amplitude increases. Essentially, Lines 2 and 3 work together to reduce non-solution amplitudes and amplify the solution amplitude.

4. Steps 2 and 3 incrementally increase the solution state's amplitude by about \(O(1/\sqrt{2^n})\) each iteration. Over \(\lfloor\frac{\pi}{4}\sqrt{2^n}\rfloor\) iterations, this amplification drives the probability close to 1, so measurement collapses the state to the solution. For \(M\) valid solutions, typically \(\left\lfloor\frac{\pi}{4}\sqrt{2^n/M}\right\rfloor\) iterations suffice to find the solution. If \(M\) is unknown, it can be estimated using the quantum counting algorithm~\cite{brassard1998quantum}.

 
\textbf{Intuition.}
To apply Grover's search to MKP, consider a graph with \(n\) vertices, yielding \(2^n\) possible subgraphs. The goal is to find a maximum \(k\)-plex among these subgraphs, which is similar to searching for a valid solution among \(2^n\) candidates. The challenge is developing a unique oracle to identify and invert the amplitude of the desired state. We illustrate the construction of this specialized oracle using quantum circuit. Due to the general applicability of the diffusion operator and space constraints, its details are omitted.




\section{Gate-Based Quantum Algorithms for MKP}\label{sec:3}
Given a $G(V, E)$, the naive method to identify the largest $k$-plex is to examine all vertex subsets, assessing each for $k$-plex formation, and noting the largest one. With the graph comprising $2^n$ potential subgraphs, this approach has a time complexity of $O^*(2^n)$. 
To tackle this efficiently, Grover's search offers a potential solution.  
We introduce two algorithms: qTKP, for locating a $k$-plex with a minimum size $T$, and qMKP, which utilizes binary search in conjunction with qTKP to determine the maximum $k$-plex.

\begin{figure}[t]
\vspace{-5pt}

	\centering
	{\includegraphics[width=6cm]{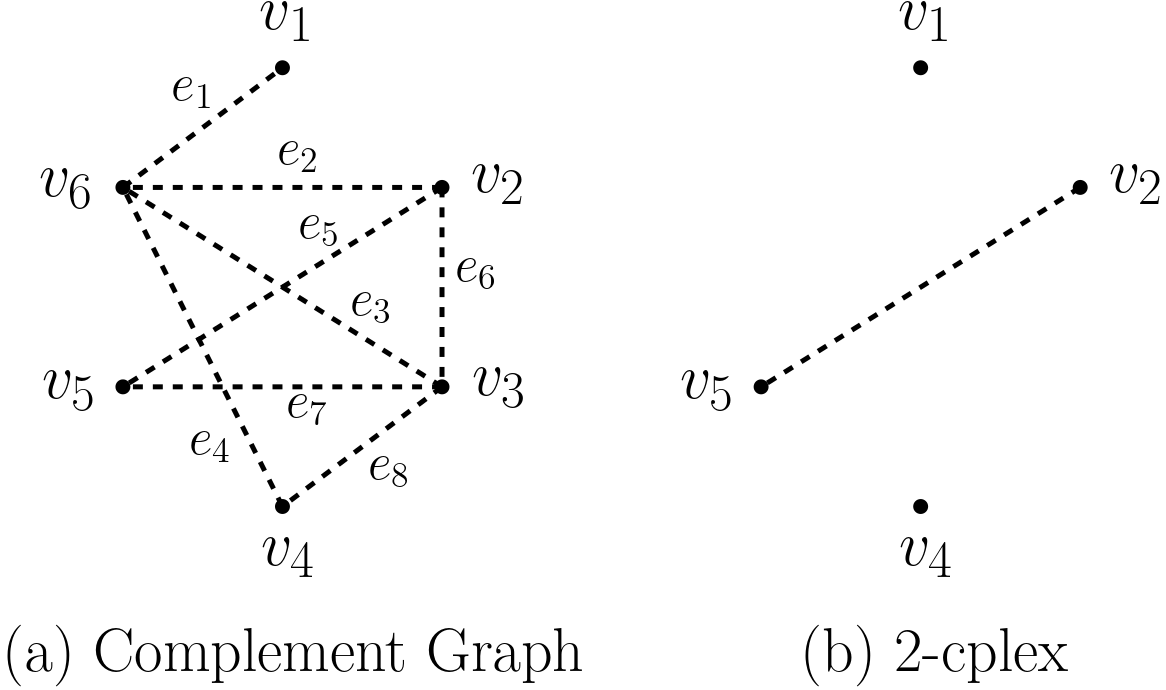}}

	\vspace{-6pt}
	\captionsetup{justification=centering}
	\caption{The complementary graph and a $2$-cplex}\label{cgraph}
	\vspace{-12pt}
\end{figure}

\subsection{The Ideas of qTKP: Quantum Algorithm to Find a $k$-Plex with a Minimum Size $T$}

To represent $n$ vertices in a graph with $n$ qubits, we use one-hot encoding. Each qubit stands for a vertex and the binary digit  $1$ (included) or $0$ (excluded) signifies  whether a vertex is part of a subset. An $n$-qubit state then represents a specific vertex combination.
For clarity, we use the graph depicted in Figure~\ref{graph} as an example, which has vertices $V = \{v_1,v_2,v_3,v_4,v_5,v_6\}$. A vertex subset, such as $\{v_1,v_4\}$, is encoded as the quantum state $\ket{100100}$, or simply $\ket{36}$.
Now we introduce the notion of a complementary graph to reframe the MKP in a different context. The reasoning for this strategic approach will be explained in subsequent discussions.

%
%
%

\begin{definition}[\textbf{Complementary Graph}]
	For a graph $G(V,E)$, its complementary graph $\overline{G}$ is $\overline{G}(V, V\times V - E)$.
\end{definition}

A set of vertices forming a $k$-plex in graph $G$ is equivalent to a $k$-cplex in the complementary graph $\overline{G}$:

\begin{definition}[\textbf{$k$-Cplex}]
In a vertex set $C \subseteq V$ that forms a $k$-cplex, each vertex $v$ has a degree $d_P(v) \leq k-1$, where $k$ is an integer within $[1,|P|-1]$.
\end{definition}

In a $k$-cplex, each vertex links to at most $k-1$ other vertices. Identifying the largest $k$-plex in $G$ is equivalent to finding the largest $k$-cplex in $\overline{G}$. Figure~\ref{cgraph} displays the complementary graph for Figure~\ref{graph} and its maximum $2$-cplex.
To find the largest $k$-cplex, we confront multiple challenges:
\begin{itemize}
	\item [I.] Encoding the topology of the complementary graph $\overline{G}$ into a quantum circuit, effectively representing the graph's structure within quantum computational models.
	\item [II.] Leveraging quantum circuits to count a vertex's degree in a subgraph, aligning with the $k$-cplex definition.
	\item [III.] Utilizing quantum circuits to compare a vertex's degree with $k-1$.
	\item [IV.] Implementing a quantum circuit approach to ascertain the size of a subgraph and comparing it with a predefined threshold $T$.
\end{itemize}


To understand why we use a complementary graph for locating $k$-cplexes rather than $k$-plexes, consider Challenge III. A $k$-plex, by definition, requires each vertex to connect with at least \(|P| - k\) others in the subgraph, where \(|P|\) varies with the subgraph's size. This variability can result in negative values for \(|P| - k\), complicating quantum circuit implementation.
Addressing these challenges allows us to create a quantum circuit that discerns if a subgraph is at least a $k$-cplex of size $T$, forming the essential oracle for Grover's search. We will now outline solutions for each challenge.

\subsection{Challenge I: Graph Encoding}
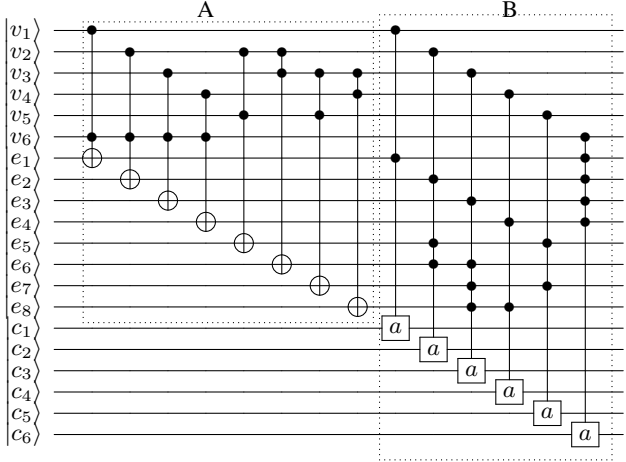
\begin{figure}[t]
\vspace{-23pt}
 \begin{small}
\[
\hspace{0.23cm}
\Qcircuit @C=.45em @R=-.25em @! {
& & & & &\mbox{A} & & &  & & & & &\hspace{0.02cm}\mbox{B} & & & & &\\
&\lstick{\ket{v_1}}   & \ctrl{5} & \qw & \qw & \qw & \qw & \qw & \qw & \qw & \ctrl{6} & \qw & \qw & \qw & \qw & \qw & \qw\\
&\lstick{\ket{v_2}}   & \qw & \ctrl{4} & \qw & \qw & \ctrl{3} & \ctrl{1} & \qw & \qw & \qw & \ctrl{6} & \qw & \qw & \qw & \qw & \qw\\
&\lstick{\ket{v_3}}   & \qw & \qw & \ctrl{3} & \qw & \qw & \ctrl{9} & \ctrl{2} & \ctrl{1} & \qw & \qw & \ctrl{6} & \qw & \qw & \qw & \qw\\
&\lstick{\ket{v_4}}   & \qw & \qw & \qw & \ctrl{2} & \qw & \qw & \qw & \ctrl{10} & \qw & \qw & \qw & \ctrl{6} & \qw & \qw & \qw\\
&\lstick{\ket{v_5}}   & \qw & \qw & \qw & \qw & \ctrl{6} & \qw & \ctrl{8} & \qw & \qw & \qw & \qw & \qw & \ctrl{6} & \qw & \qw\\
&\lstick{\ket{v_6}}   & \ctrl{1} & \ctrl{2} & \ctrl{3} &\ctrl{4} & \qw & \qw & \qw & \qw & \qw & \qw & \qw & \qw & \qw & \ctrl{1} & \qw\\
&\lstick{\ket{e_1}}   & \targ & \qw & \qw & \qw & \qw & \qw & \qw & \qw & \ctrl{8} & \qw & \qw & \qw & \qw & \ctrl{1} & \qw\\
&\lstick{\ket{e_2}}   & \qw & \targ & \qw & \qw & \qw & \qw & \qw & \qw & \qw & \ctrl{3} & \qw & \qw & \qw & \ctrl{1} & \qw\\
&\lstick{\ket{e_3}}   & \qw & \qw & \targ  & \qw & \qw & \qw & \qw & \qw & \qw & \qw & \ctrl{3} & \qw & \qw & \ctrl{1} & \qw\\
&\lstick{\ket{e_4}}   & \qw & \qw & \qw & \targ & \qw & \qw & \qw & \qw & \qw & \qw & \qw & \ctrl{4} & \qw & \ctrl{10} & \qw\\
&\lstick{\ket{e_5}}   & \qw & \qw & \qw & \qw & \targ & \qw & \qw & \qw & \qw & \ctrl{1} & \qw & \qw & \ctrl{2} & \qw & \qw\\
&\lstick{\ket{e_6}}   & \qw & \qw & \qw & \qw & \qw & \targ & \qw & \qw & \qw & \ctrl{4} & \ctrl{1} & \qw & \qw & \qw & \qw\\
&\lstick{\ket{e_7}}   & \qw & \qw & \qw & \qw & \qw & \qw & \targ & \qw & \qw & \qw & \ctrl{1} & \qw & \ctrl{6} & \qw & \qw\\
&\lstick{\ket{e_8}}   & \qw & \qw & \qw & \qw & \qw & \qw & \qw & \targ & \qw & \qw & \ctrl{3} & \ctrl{4} & \qw & \qw & \qw\\
&\lstick{\ket{c_1}}   & \qw & \qw & \qw & \qw & \qw & \qw & \qw & \qw & \gate{a} & \qw & \qw & \qw & \qw & \qw & \qw\\
&\lstick{\ket{c_2}}   & \qw & \qw & \qw & \qw & \qw & \qw & \qw & \qw & \qw & \gate{a} & \qw & \qw & \qw & \qw & \qw\\
&\lstick{\ket{c_3}}   & \qw & \qw & \qw & \qw & \qw & \qw & \qw & \qw & \qw & \qw & \gate{a} & \qw & \qw & \qw & \qw\\
&\lstick{\ket{c_4}}   & \qw & \qw & \qw & \qw & \qw & \qw & \qw & \qw & \qw & \qw & \qw & \gate{a}  & \qw & \qw & \qw\\
&\lstick{\ket{c_5}}   & \qw & \qw & \qw & \qw & \qw & \qw & \qw & \qw & \qw & \qw & \qw & \qw & \gate{a} & \qw & \qw\\
&\lstick{\ket{c_6}}   & \qw & \qw & \qw & \qw & \qw & \qw & \qw & \qw & \qw & \qw & \qw & \qw & \qw & \gate{a} & \qw
\gategroup{2}{3}{15}{10}{.5em}{.}
\gategroup{2}{11}{21}{16}{1.1em}{.}
}
\]
 \end{small}
\vspace{-10pt}
\caption{Quantum circuit of complementary graph encoding and edge count. \textbf{A}: to encode the complementary graph into   circuit; \textbf{B}: to count how many edges $v_i$ has and then store this number into $c_i$.}
\label{qcencoding}
\vspace{-10pt}
\end{figure}

To encode a complementary graph's topology into a quantum circuit, we represent its $n$ vertices with $n$ qubits, labeled as $\{\ket{v_i}\}$. We also use additional $(n^2-m)$ auxiliary qubits, $\{\ket{e_i}\}$,  to represent the edges in the complementary graph, initializing them to $\ket{0}$.
For an edge $e_k$ connecting vertices $v_i$ and $v_j$, a C$^2$NOT gate is used to link $\ket{v_i}$, $\ket{v_j}$, and $\ket{e_k}$, with $\ket{e_k}$ being the target. This means the edge $e_k$ is activated to $\ket{1}$ by the C$^2$NOT gate only if both $v_i$ and $v_j$ are present in a subgraph.
Considering a base state $\ket{x}$, with $x$ in the range from $0$ to $2^n-1$, the C$^2$NOT gates activate all corresponding edge qubits to $\ket{1}$. For illustration, refer to Figure~\ref{cgraph} and the dashed box marked \textit{A} in Figure~\ref{qcencoding}.
For example, for $\ket{x} = \ket{33} = \ket{100001} = \ket{{v_1,v_6}}$, only edge $\ket{e_1}$ is activated to $\ket{1}$, while the others remain $\ket{0}$. Through this approach, the graph's topological structure is effectively encoded into a quantum circuit.


\textbf{Graph Encoding Complexity Analysis.}
In quantum algorithms, the space complexity is measured by the number of qubits used in the quantum circuit. Time complexity, on the other hand, is quantified by the number of quantum gates utilized.
Given a graph $G(V,E)$ with $n$ vertices and $m$ edges, we see that the number of qubits utilized to encode the graph is $n+n^2-m = O(n^2)$. 
The number of C$^2$NOT gates is $O(n^2)$.

\subsection{Challenge II: Degree Counting (Oracle Part 1)}

\textbf{The High-Level Idea.}
In a subgraph, it's necessary to count the degree of each vertex. For a specific vertex $v_i$, this means summing the number of edges connected to it within the subgraph and storing this count in a set of auxiliary qubits $\{\ket{c_i}\}$. Notably, the edges connected to $v_i$ within the subgraph are already activated to $\ket{1}$ during the graph encoding process, while edges connected to $v_i$ outside the subgraph remain in their initial state $\ket{0}$. Therefore, summing all the edges associated with $v_i$ in the full graph effectively filters out those outside the subgraph. 
Figure~\ref{qcencoding} illustrates this process in box B. For clarity, the Control-$a$ gate is a conceptual quantum gate representing the summation of the states of all edges connected to a vertex $\ket{v_i}$ appearing in the subgraph, with the result stored in $\ket{c_i}$. Next, we will discuss how to implement this abstract quantum gate, Control-$a$.
Assuming for vertex $v_3$, the four edges $\{e_3,e_6,e_7,e_8\}$ it is  connected to are activated in the current subgraph with states $\ket{e_3}=\ket{e_6}=\ket{e_7}=\ket{1}$, and $\ket{e_8}=\ket{0}$. The challenge is to implement a quantum circuit that maps this edge activation to a count, transforming $\ket{e_3e_6e_7e_8}=\ket{1110}$ into $\ket{3}$, and storing this result in $\ket{c_3}$. In other words, how can a quantum circuit count the number of 1s in a binary string? The first step is  to construct a quantum circuit for addition of two one-bit numbers with carry. The second step builds upon this to implement addition for two multi-bit numbers. Finally, this results in a degree count quantum circuit capable of calculating the sum of activated edges.


\begin{figure}[t]
\vspace{-22pt}
\begin{small}
\[
\hspace*{-1cm}
\Qcircuit @C=2em @R=0.2em @! {
&&\mbox{A}&\mbox{B}&\mbox{C}&\mbox{D}&\mbox{E}&\\
&\lstick{\ket{x}} & \ctrl{1} & \ctrl{1} & \qw & \qw & \qw & \qw\\
&\lstick{\ket{y}} & \ctrl{2} & \targ & \ctrl{1} & \ctrl{1} & \qw & \qw\\
&\lstick{\ket{C_{in}}} & \qw & \qw & \ctrl{2} & \targ & \qw & \rstick{\ket{sum}}\qw\\
&\lstick{\ket{0}} & \targ  & \qw & \qw & \qw & \ctrl{1} & \qw\\
&\lstick{\ket{0}} & \qw & \qw & \targ & \qw & \targ & \rstick{\ket{C_{out}}}\qw
\gategroup{2}{3}{5}{3}{.8em}{.}
\gategroup{2}{4}{3}{4}{.8em}{.}
\gategroup{3}{5}{6}{5}{.8em}{.}
\gategroup{3}{6}{4}{6}{.8em}{.}
\gategroup{5}{7}{6}{7}{.8em}{.}
}
\]
\end{small}
\vspace{-12pt}
\caption{Quantum circuit of integer addition. \textbf{A}: to compute \ket{x\wedge y}; \textbf{B}: to compute \ket{x\oplus y}; \textbf{C}: to compute \ket{C_{in}\wedge (x\oplus y)}; \textbf{D}: to compute \ket{sum} = \ket{x\oplus y\oplus C_{in}}; \textbf{E}: to compute \ket{C_{out}} = \ket{[x\wedge y] \oplus [C_{in}\wedge (x\oplus y)]}.  } 
\label{qcadd}
\vspace{-8pt}
\end{figure}
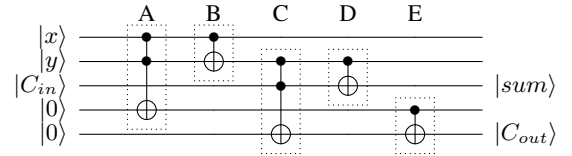


\begin{figure}[t]
\begin{small}
\[
\hspace*{-2cm}
\Qcircuit @C=.1em @R=-1.5em @! {
&\lstick{\ket{0}}  & \multigate{2}{add} & \qw & \qw & \qw & \qw & \qw\\
&\lstick{\ket{x_3}}  & \ghost{add}  & \qw & \qw & \qw  & \qw & \rstick{\ket{x_3\oplus y_3}}\qw\\
&\lstick{\ket{y_3}}  & \ghost{add}  & \qw & \multigate{2}{add} & \qw & \qw & \rstick{\ket{C_3}}\qw\\
&\lstick{\ket{x_2}}  & \qw  & \qw & \ghost{add} & \qw  & \qw & \rstick{\ket{x_2\oplus y_2 \oplus C_3}}\qw\\
&\lstick{\ket{y_2}}  & \qw  & \qw & \ghost{add} & \qw & \multigate{2}{add} & \rstick{\ket{C_2}}\qw\\
&\lstick{\ket{x_1}}  & \qw  & \qw & \qw & \qw & \ghost{add}   & \rstick{\ket{x_1\oplus y_1\oplus C_2}}\qw\\
&\lstick{\ket{y_1}}  & \qw  & \qw & \qw & \qw & \ghost{add} & \rstick{\ket{C_1}}\qw\\
}
\]
\end{small}
\vspace{-9pt}
\caption{Simplified version (auxiliary qubits are omitted) of a multi-qubit quantum addition circuit, showing $\ket{x_1x_2x_3 + y_1y_2y_3}$ as an example. The quantum gates labeled {\it add} correspond to the   circuit depicted in Figure~\ref{qcadd}. 
  $C_i$ represents the carry digit.} 
\label{qcmadd}
\vspace{-10pt}
\end{figure}
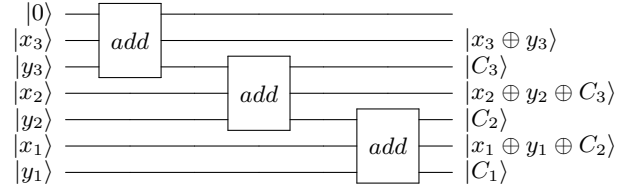

\textbf{One-Qubit Addition.}
Suppose we aim to compute the sum of two qubits $\ket{x}$ and $\ket{y}$, while also considering a carry-in qubit $\ket{C_{in}}$. The result is to be stored in $\ket{sum}$ and the carry-out qubit $\ket{C_{out}}$. We utilize the following construction equation for this purpose:
\begin{equation}\label{eq:add}
	\begin{aligned}
		\ket{sum} &= \ket{x\oplus y \oplus C_{in}}\\
		\ket{C_{out}} &= \ket{[x\wedge y] \oplus [C_{in}\wedge (x\oplus y)]} 
	\end{aligned}
\end{equation}
To construct a quantum circuit based on this equation, we require two auxiliary qubits for storing intermediate results. The designed quantum circuit is shown in Figure~\ref{qcadd}. The five boxes labeled A-E in the figure correspond to the computation of different terms in Eq.~\ref{eq:add}. It's important to note that the logical operations $\wedge$ and $\oplus$ used in Eq.~\ref{eq:add} are replaced by CNOT and C$^2$NOT gates, respectively, in the quantum circuit.
The number of qubits to implement the one-qubit addition is $5 = O(1)$, and the number of gates is $5 = O(1)$.

\textbf{Multi-Qubit Addition.}
However, when dealing with more than two qubits, it is  necessary to consider multi-qubit quantum addition. Specifically, given two multi-qubit $\ket{x=x_1x_2...x_s}$ and $\ket{y=y_1y_2...y_s}$, the task is to compute $\ket{x+y}$. This can be achieved by sequentially calculating $\ket{x_i+y_i}$ for each $i$ from $s$ to 1, while accounting for the carry. Figure~\ref{qcmadd} demonstrates the construction of a multi-qubit addition quantum circuit, using $s=3$ as an example. In this figure, one-qubit addition is abstractly represented by a quantum gate labeled {\it add}. For simplicity, auxiliary bits are not shown in the diagram. 
For the addition of two $s$-qubit numbers, it is necessary to invoke one-qubit addition $s$ times. The required number of qubits is $5\times s = O(s)$, and the number of gates needed is also $5\times s = O(s)$.

\textbf{Degree Counting Complexity Analysis.}
We have constructed a quantum circuit capable of counting the degrees of vertices within a subgraph. 
To count the degrees for up to $n$ vertices, where each vertex's degree could be at most $n-1$, we need to perform at most $(n-1)$ additions using $\log (n-1)$-length multi-qubit addition for each vertex. Therefore, the required number of qubits is $n \times  (n-1) \times 5\log (n-1) = O(n^2\log n)$, and the number of gates needed is also $n \times  (n-1) \times 5\log (n-1) = O(n^2\log n)$.

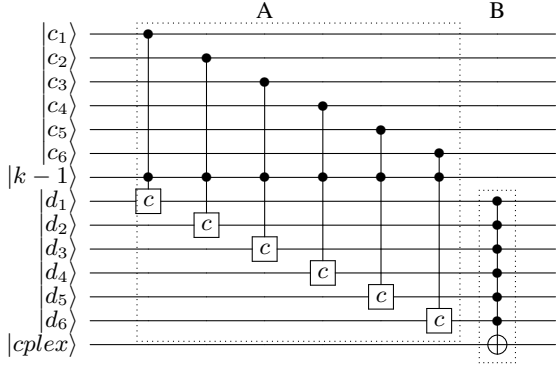
\begin{figure}[t]
\vspace{-20pt}
 \begin{small}
\[
\hspace{0.5cm}
\Qcircuit @C=1.4em @R=-.04em @! {
& & & & \mbox{A}& & & &\mbox{B} & &  \\
&\lstick{\ket{c_1}} & \ctrl{6} & \qw  & \qw & \qw & \qw & \qw & \qw & \qw\\
&\lstick{\ket{c_2}} & \qw & \ctrl{5}  & \qw & \qw & \qw & \qw & \qw & \qw\\
&\lstick{\ket{c_3}} & \qw & \qw  & \ctrl{4} & \qw & \qw & \qw & \qw & \qw\\
&\lstick{\ket{c_4}} & \qw & \qw  & \qw & \ctrl{3} & \qw & \qw & \qw & \qw\\
&\lstick{\ket{c_5}} & \qw & \qw  & \qw & \qw & \ctrl{2} & \qw & \qw & \qw\\
&\lstick{\ket{c_6}} & \qw & \qw  & \qw & \qw & \qw & \ctrl{1} & \qw & \qw\\
&\lstick{\ket{k-1}} & \ctrl{1} & \ctrl{2}  & \ctrl{3} & \ctrl{4} & \ctrl{5} & \ctrl{6} & \qw & \qw\\
&\lstick{\ket{d_1}} & \gate{c} & \qw  & \qw & \qw & \qw & \qw & \ctrl{1} & \qw\\
&\lstick{\ket{d_2}} & \qw & \gate{c}  & \qw & \qw & \qw & \qw & \ctrl{1} & \qw\\
&\lstick{\ket{d_3}} & \qw & \qw  & \gate{c} & \qw & \qw & \qw & \ctrl{1} & \qw\\
&\lstick{\ket{d_4}} & \qw & \qw  & \qw & \gate{c} & \qw & \qw & \ctrl{1} & \qw\\
&\lstick{\ket{d_5}} & \qw & \qw  & \qw & \qw & \gate{c} & \qw & \ctrl{1} & \qw\\
&\lstick{\ket{d_6}} & \qw & \qw  & \qw & \qw & \qw & \gate{c} & \ctrl{1} & \qw\\
&\lstick{\ket{cplex}} & \qw & \qw  & \qw & \qw & \qw & \qw & \targ & \qw
\gategroup{2}{3}{14}{8}{.7em}{.}
\gategroup{9}{9}{15}{9}{.7em}{.}
}
\]
 \end{small}
\vspace{-12pt}
\caption{Quantum circuit of edge comparison and cplex check. \textbf{A}: for each vertex $v_i$, to compare its edge number with the threshold $k-1$; \textbf{B}: to check whether the subgraph is a $k$-cplex.}
\label{qcplex}
\vspace{-10pt}
\end{figure}

\subsection{Challenge III: Degree Comparison (Oracle Part 2)}

\textbf{High-Level Idea.}
We have now recorded the degree of each vertex $v_i$  within $\ket{c_i}$. (Note that each $\ket{c_i}$ actually represents a set of auxiliary bits. To simplify the representation of the quantum circuit, only a single wire is used to depict this in Figures~\ref{qcencoding} and \ref{qcplex}.) Next, to adhere to the $k$-cplex definition, we compare each vertex's degree with $k-1$. Figure~\ref{qcplex} uses a conceptual control-$c$ gate for this comparison, storing the result in the new auxiliary bits $\ket{d_i}$. If $c_i < k-1$, then $\ket{d_i}$ is set to $\ket{1}$; otherwise, it is  $\ket{0}$. This is depicted in box A of Figure~\ref{qcplex}.
A complementary subgraph qualifies as a $k$-cplex (and thus, the original subgraph as a $k$-plex) only if all $\ket{d_i}$ are $\ket{1}$ for each $i$ in the range [1, n]. The validation of this condition is illustrated in box B of Figure~\ref{qcplex}.

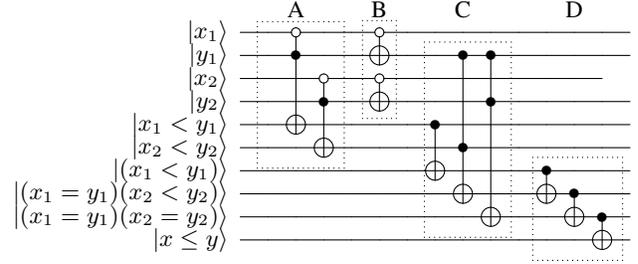
\begin{figure}[t]
\vspace{-20pt}
\begin{small}
\[
\hspace*{2cm}
\Qcircuit @C=.35em @R=0.15em @! {
&&&\mbox{A}&&&\mbox{B}&&&\mbox{C}&&&&\mbox{D}&\\
&\lstick{\ket{x_1}} & \qw & \ctrlo{1} & \qw & \qw & \ctrlo{1}  & \qw& \qw & \qw & \qw & \qw & \qw & \qw & \qw & \qw\\
&\lstick{\ket{y_1}} & \qw & \ctrl{3} & \qw & \qw & \targ  & \qw& \qw & \ctrl{4} & \ctrl{2} & \qw & \qw & \qw & \qw & \qw\\
&\lstick{\ket{x_2}} & \qw & \qw & \ctrlo{1} & \qw & \ctrlo{1}  & \qw & \qw & \qw & \qw & \qw & \qw & \qw & \qw\\
&\lstick{\ket{y_2}} & \qw & \qw & \ctrl{2} & \qw & \targ & \qw & \qw & \qw & \ctrl{5} & \qw & \qw & \qw & \qw & \qw\\
&\lstick{\ket{x_1<y_1}} & \qw & \targ & \qw & \qw & \qw & \qw & \ctrl{2} & \qw & \qw & \qw & \qw & \qw & \qw & \qw \\
&\lstick{\ket{x_2<y_2}} & \qw & \qw & \targ  & \qw& \qw & \qw & \qw & \ctrl{2} & \qw & \qw & \qw & \qw & \qw & \qw \\
&\lstick{\ket{(x_1<y_1)}} & \qw & \qw & \qw  & \qw& \qw & \qw & \targ & \qw & \qw & \qw & \ctrl{1} & \qw & \qw & \qw \\
&\lstick{\ket{(x_1=y_1)(x_2<y_2)}} & \qw & \qw & \qw & \qw & \qw & \qw & \qw &\targ & \qw & \qw & \targ & \ctrl{1} & \qw & \qw \\
&\lstick{\ket{(x_1=y_1)(x_2=y_2)}} & \qw & \qw & \qw & \qw & \qw & \qw & \qw & \qw & \targ & \qw & \qw & \targ  & \ctrl{1} & \qw\\
&\lstick{\ket{x\leq y}} & \qw & \qw & \qw & \qw & \qw & \qw & \qw & \qw & \qw & \qw & \qw & \qw  & \targ & \qw 
\gategroup{2}{3}{7}{5}{.9em}{.}
\gategroup{2}{7}{5}{7}{.6em}{.}
\gategroup{3}{9}{10}{11}{.9em}{.}
\gategroup{8}{13}{11}{15}{.9em}{.}
}
\]
\end{small}
\vspace{-12pt}
\caption{Quantum circuit of integer comparison. {\textbf{A}}: to compute $\ket{x_i<y_i}$; {\textbf{B}}: to compute $\ket{x_i=y_i}$; \textbf{C}: to compute each item in the R.H.S. of the Eq.~\ref{eq:comp}; \textbf{D}: to compute \ket{x\leq y}.} 
\label{qccomp}
\vspace{-12pt}
\end{figure}

\textbf{Degree Comparison.}
To implement the control-$c$ gate, which compares the degree with $k-1$, we simplify the problem to comparing two non-negative integers $x$ and $y$. Each integer is represented in binary form as $x = x_1x_2x_3...x_s$ and $y = y_1y_2y_3...y_s$, where $s$ is their binary bit length. The comparison starts with the most significant bit (the leftmost bit). If the bits are different, the comparison ends; otherwise, it continues to the next bit. This process is repeated until a difference is found or all bits are compared. This logic is encapsulated in a mathematical expression: 
\begin{equation}\label{eq:comp}
	\begin{aligned}
		x\leq y \Leftrightarrow [x_1 < y_1] & \vee [(x_1=y_1) (x_2<y_2)]\\
		& \vee [(x_1=y_1)\wedge(x_2=y_2) \wedge (x_3<y_3)] \vee ...\\
		&  \vee [(x_1=y_1)\wedge(x_2=y_2)\wedge...\wedge(x_s=y_s)]
	\end{aligned}
\end{equation}
For each one-bit comparison, we have:
\begin{equation}\label{eq:1comp}
	\begin{aligned}
		x_i < y_i &\Leftrightarrow \overline{x_i} \wedge y_i\\
		x_i = y_i &\Leftrightarrow \overline{x_i} \oplus y_i
	\end{aligned}
\end{equation}
From Eq.~\ref{eq:comp} and Eq.~\ref{eq:1comp}, we can construct a quantum circuit for degree comparison. 
Taking $s=2$ as an example, we illustrate this quantum circuit in Figure~\ref{qccomp}. The circuit is divided into four boxes, A-D. Box A employs the logic of $\overline{x_i} \wedge y_i$ to determine if $x_i < y_i$. Box B uses the logic of $\overline{x_i} \oplus y_i$ to assess if $x_i = y_i$. Box C constructs each discriminator following the $\vee$ in the right-hand side of Eq.~\ref{eq:comp}. Finally, box D uses the $\vee$ logic to combine the discriminators formed in box C to evaluate the final  condition $x_1x_2 \leq y_1y_2$.
Using this degree comparison circuit, we can implement the control-$c$ gate in Figure~\ref{qcplex}. 
For comparing $s$-bit numbers, the degree comparison circuit requires $s + s + 1 = O(s)$ auxiliary qubits and needs $s + s + s + 1 + s = O(s)$ gates. 

\textbf{Degree Comparison Complexity Analysis.}
Given that the maximum degree of each vertex is $n^2 - 1$, the maximum value for $s$ could be $\log (n^2 - 1)$. Thus, each vertex's degree comparison requires $O(\log (n^2 - 1)) = O(\log n)$ auxiliary bits and $O(\log (n^2 - 1)) = O(\log n)$ gates. As shown in Figure~\ref{qcplex}, a maximum of $n$ comparisons may be needed, leading to a total qubit number  $O(n\log n)$ and a total gate number  $O(n\log n)$.

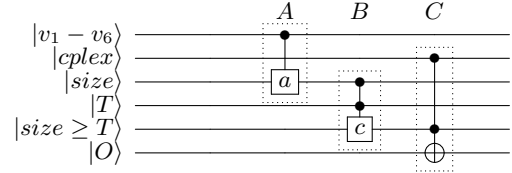
\begin{figure}[t]
 \begin{small}
\[
\Qcircuit @C=2.0em @R=-.15em @! {
& & &A &B &C & \\
&\lstick{\ket{v_1-v_6}} & \qw & \ctrl{2}  & \qw\ & \qw & \qw\\
&\lstick{\ket{cplex}} & \qw & \qw  & \qw & \ctrl{3} & \qw\\
&\lstick{\ket{size}} & \qw & \gate{a}  & \ctrl{1} & \qw & \qw\\
&\lstick{\ket{T}} & \qw & \qw  & \ctrl{1} & \qw & \qw\\
&\lstick{\ket{size\geq T}} & \qw & \qw  & \gate{c} & \ctrl{1} & \qw\\
&\lstick{\ket{O}} & \qw & \qw  & \qw & \targ & \qw
\gategroup{2}{4}{4}{4}{.7em}{.}
\gategroup{4}{5}{6}{5}{.7em}{.}
\gategroup{3}{6}{7}{6}{.7em}{.}
}
\]
 \end{small}
\vspace{-12pt}
\caption{Quantum circuit of plex size check. \textbf{A}: to determine the size of the subgraph by counting how many vertices it has; \textbf{B}: to compare the size with the given threshold $T$; \textbf{C}: to mark the solution state by the oracle qubit $\ket{O}$. }
\label{qcoracle}
\vspace{-10pt}
\end{figure}

\subsection{Challenge IV: Size Determinition (Oracle Part 3)}
We have now developed a quantum circuit to identify whether a  subgraph is a $k$-cplex. The next step is to determine if this subgraph's size meets or exceeds a specific threshold $T$. This is achieved by extending our earlier processes of degree counting and comparison.
In this final phase, depicted in Figure~\ref{qcoracle}, we use:
1) The control-$a$ gate (from the degree counting circuit) to count the number of vertices, storing this count in auxiliary bits labeled $\ket{size}$ (refer to box A).
2) The control-$c$ gate (from the degree comparison circuit) to compare this counted size with the threshold $T$ (see box B).
The culmination of this process involves the oracle qubit $\ket{O}$. This qubit is flipped to $\ket{1}$, indicating $\ket{cplex} = \ket{size \geq T} = \ket{1}$ if the subgraph's size meets the threshold (as shown in box C). Through these steps, we've successfully designed a comprehensive oracle quantum circuit for locating a $k$-cplex of at least size $T$ in a complementary graph.


\textbf{Size Determinition Complexity Analysis.}
In Figure~\ref{qcoracle}, box A is used  to sum up vertices $n$ times, with the sum potentially having a maximum bit length of $\log n$. Therefore, it requires at most $O(n\log n)$ qubits and $O(n\log n)$ quantum gates. Box B involves a single comparison of numbers with up to $\log n$ bits, necessitating at most $O(\log n)$ qubits and $O(\log n)$ quantum gates. Box C simply includes a basic quantum gate. Hence, the total space complexity of Figure~\ref{qcoracle} is $O(n\log n) + O(\log n) = O(n\log n)$, and the total time complexity is similarly $O(n\log n) + O(\log n) = O(n\log n)$.

\textbf{Full Oracle Complexity Analysis.}
Finally, by summing up the space and time complexities of the graph encoding (Figure~\ref{qcencoding}) and the three components of the complete oracle - degree counting (Figure~\ref{qcencoding}), degree comparison (Figure~\ref{qcplex}), and size determination (Figure~\ref{qcoracle}) - we arrive at the overall complexity of the oracle:

Space complexity: 
\begin{equation*}
	O(n^2)+O(n^2\log n)+O(n\log n)+O(n\log n)=O(n^2\log n)
\end{equation*}

Time complexity:
\begin{equation*}
	O(n^2)+O(n^2\log n)+O(n\log n)+O(n\log n)=O(n^2\log n)
\end{equation*}

\begin{algorithm}[tttt]
        \caption{Quantum $k$-Plex with Size $T$ Search (qTKP)}
        \begin{algorithmic}[1] 
            \Require Graph $G(V,E)$, $k$, size threshold $T$;
            \Ensure A $k$-plex with size no less than $T$ or $\emptyset$;
            \State Convert $G$ into the complementary graph $\overline G$;
            \State Prepare the initial state to be an equal superposition of $2^n$ possible subsets of $V$;
            \State Use the oracle  with $\ket{O}$ to flip  the amplitude signs of the $k$-cplexes with size $T$;
            \State Use a diffusion operator to inverse the amplitude of each base state about the amplitude average; 
            \State Repeat Line 2\&3 for $\lfloor\frac{\pi}{4}\sqrt{2^n/M}\rfloor$ times, then measure  the $n$ vertex qubits;
			\State Output the $k$-cplex ($k$-plex) or $\emptyset$. 
        \end{algorithmic}
    \label{algo:qTKP} 
   
    \end{algorithm}

\subsection{Our Algorithm: qTKP}

We have now assembled an oracle for identifying if a subgraph in a complementary graph qualifies as a $k$-cplex of size at least $T$. This prepares us to complete the qTKP algorithm.
In Grover's search (Step 2), our goal is to invert the amplitude of a solution state $\ket{x}$. To achieve this, we set the oracle qubit $\ket{O}$ to $\ket{1}$ and then apply an $H$ gate to transition it to $(\ket{0}-\ket{1})/\sqrt{2}$. Thus, the system state becomes $\ket{x}\ket{O} = \ket{x}(\ket{0}-\ket{1})/\sqrt{2}$.
As per Figure~\ref{qcoracle}, when a solution state $\ket{x}$ is processed by box C, $\ket{O}$ gains a negative sign. This sign is then distributed across the tensor product, changing $\ket{x}\ket{O}$ to $-\ket{x}\ket{O}$. Consequently, the amplitude of $\ket{x}$ is inverted.
With the oracle in place, we can now incorporate it into Grover's search to develop the qTKP algorithm, detailed in Algorithm~\ref{algo:qTKP}.

\begin{figure*}[t]
 \begin{small}
\vspace{-22pt}
\[
\Qcircuit @C=.8em @R=-2.53em @! {
& &\mbox{A} & & &\mbox{B} & & &\mbox{C} & &\hspace{-45pt}\mbox{repeat}  &\mbox{measure}\\
&\lstick{\ket{v_1}} & \gate{H} & \qw & \multigate{8}{U_{check}} & \qw & \multigate{8}{U_{check}^\dag} & \qw & \multigate{5}{U_{Diff}} & \qw &\hspace{-45pt}... & \meter \\
&\lstick{\ket{v_2}} & \gate{H} & \qw & \ghost{U_{check}} & \qw & \ghost{U_{check}^\dag} & \qw & \ghost{U_{Diff}} & \qw &\hspace{-45pt}... & \meter\\
&\lstick{\ket{v_3}} & \gate{H} & \qw & \ghost{U_{check}} & \qw & \ghost{U_{check}^\dag} & \qw & \ghost{U_{Diff}} & \qw &\hspace{-45pt}... & \meter \\
&\lstick{\ket{v_4}} & \gate{H} & \qw & \ghost{U_{check}} & \ustick{\tiny\mbox{}}\qw & \ghost{U_{check}^\dag} & \qw & \ghost{U_{Diff}} & \qw &\hspace{-45pt}... & \meter \\
&\lstick{\ket{v_5}} & \gate{H} & \qw & \ghost{U_{check}} & \qw & \ghost{U_{check}^\dag} & \qw & \ghost{U_{Diff}} & \qw &\hspace{-45pt}...  & \meter\\
&\lstick{\ket{v_6}} & \gate{H} & \qw & \ghost{U_{check}} & \qw & \ghost{U_{check}^\dag} & \qw & \ghost{U_{Diff}} & \qw &\hspace{-45pt}... & \meter \\
&\lstick{\ket{aux}} & \qw & \qw & \ghost{U_{check}} & \ustick{\tiny\mbox{sign flipping}}\qw & \ghost{U_{check}^\dag} & \qw & \qw & \qw &\hspace{-45pt}... & \qw \\
&\lstick{\ket{cplex}} & \qw & \qw & \ghost{U_{check}} & \ctrl{1} & \ghost{U_{check}^\dag} & \qw & \qw & \qw &\hspace{-45pt}... & \qw \\
&\lstick{\ket{size\geq T}} & \qw & \qw & \ghost{U_{check}} & \ctrl{1}  & \ghost{U_{check}^\dag} & \qw & \qw & \qw &\hspace{-45pt}... & \qw \\
&\lstick{\ket{O} = \ket{1}} & \gate{H} & \qw & \qw &  \targ  & \qw & \qw & \qw & \qw &\hspace{-45pt}...  & \qw
\gategroup{2}{3}{7}{3}{.5em}{--}
\gategroup{2}{4}{11}{8}{1.8em}{--}
\gategroup{2}{9}{7}{9}{.8em}{--}
\gategroup{9}{6}{11}{6}{1em}{.}
}
\]
 \end{small}
\vspace{-10pt}
\caption{Quantum circuit of qTKP for searching a $k$-plex of which the size is at least $T$. \textbf{A}: superposition of all possible subgraphs; \textbf{B}: oracle to mark the solutions by amplitude sign flipping; \textbf{C}: diffusion operator to reverse all the amplitudes about the average.}
\label{qcfull}
\vspace{-13pt}
\end{figure*}
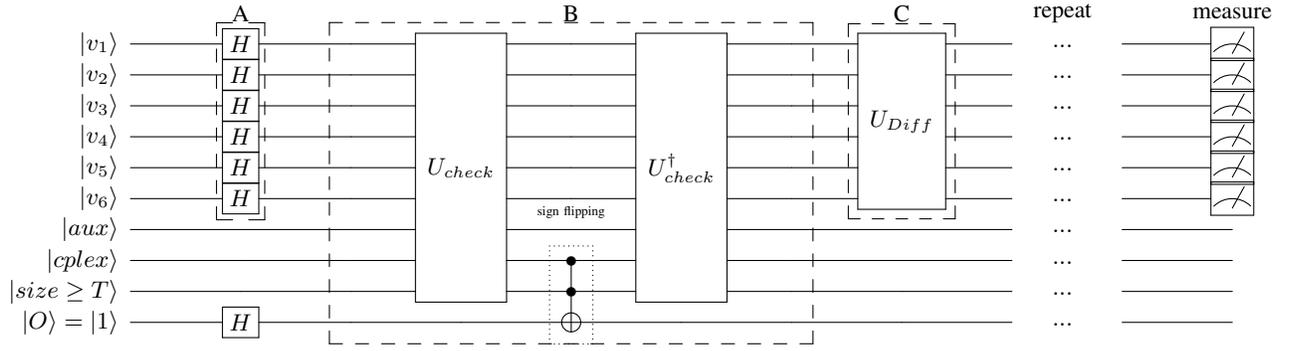

In Algorithm~\ref{algo:qTKP}, the variable $M$ represents the number of size-$T$ $k$-plexes in the graph. This number can be estimated using the quantum counting algorithm~\cite{brassard1998quantum}. The full circuit for implementing Algorithm~\ref{algo:qTKP} is depicted in Figure~\ref{qcfull}. In this representation, $\ket{aux}$ is used to collectively denote the auxiliary qubits.
The $U_{check}$ component encapsulates three key aspects of the oracle: degree counting (as shown in Figure~\ref{qcencoding}), degree comparison (Figure~\ref{qcplex}), and size determination (Figure~\ref{qcoracle}). 
To reset all auxiliary qubits to their initial states after each iteration, we apply $U_{check}^{-1}=U_{check}^\dag$ once the oracle qubit $\ket{O}$ is flipped. As the CNOT gate is its own inverse, $U^\dag$ employs the same gates as $U$, but in reverse sequence.


\textbf{qTKP Complexity Analysis.}
The qTKP algorithm requires at most $O(2^{n/2})$ oracle calls, leading to a space complexity of $O(n^2\log n)$ and a time complexity of $O(2^{n/2}n^2\log n)=O^*(2^{n/2})$.

\subsection{Our Algorithm: qMKP}
Now we can present our  algorithm qMKP to search for a maximum $k$-plex as the Quantum Maximum $k$-Plex Algorithm (shown in Algorithm~\ref{algo:qMKP}). 
The algorithm qMKP employs a binary search approach in conjunction with qTKP to determine the maximum size, denoted as $T_{MAX}$, of a $k$-plex.

\textbf{qMKP Complexity Analysis.}
The qMKP algorithm requires at most $\log n$ calls to qTKP. Consequently, the space complexity of qMKP is the same as that of qTKP, amounting to $O(n^2\log n)$. The time complexity of qMKP is $O(2^{n/2}n^2\log^2 n)=O^*(2^{n/2})$.


\begin{algorithm}[tttt]
        \caption{Quantum Maximum $k$-Plex Search  (qMKP)}
        \begin{algorithmic}[1] 
            \Require Graph $G(V,E)$, $k$;
            \Ensure A maximum $k$-plex;
            \State Use  qTKP to search for a $k$-plex with  size $T$, and find the maximum $T\in [1,|V|]$ by binary search; 
            \State Output the maximum $k$-plex. 

        \end{algorithmic}
    \label{algo:qMKP} 
   
    \end{algorithm}

\textbf{Orthogonality.}
The qMKP algorithm is orthogonal to graph reduction algorithms, such as the second-order reduction~\cite{zhou2021improving} and the core-truss co-pruning technique~\cite{chang2022efficient}. The purpose of graph reduction is to prune vertices and edges that are unlikely to be part of a maximum $k$-plex from the original graph, thereby reducing the input graph's size. This reduction should guarantee that the reduced graph still contains a maximum $k$-plex. Hence, running qMKP on a reduced graph does not affect its ability to find a solution and can help qMKP operate on slightly larger datasets within the hardware constraints of current quantum devices. Additionally, upper bounding techniques~\cite{jiang2021new} can also be integrated into the binary search process of qMKP to further enhance its efficiency.

\textbf{Progression.}
Since qMKP continuously invokes qTKP during its binary search process to output new \( k \)-plexes, it essentially operates as a progressive algorithm. The progression nature of qMKP can be analyzed theoretically. Given that qMKP calls qTKP \( O(\log n) \) times, the time taken by qMKP to output the first feasible solution is approximately the first \( O(1/\log n) \) of the total running time. It means that as the graph size increases, qMKP can obtain the first feasible solution more quickly. Furthermore, based on the characteristics of binary search, it can be inferred that the size of the first feasible solution is at least half of the optimal solution.

\textbf{Adaptability.}
The graph encoding circuit and the three-component oracle hold potential for adapting to various clique relaxation models such as  \(n\)-clan and \(n\)-club. 
An \(n\)-clan limits the longest path between any two vertices to \(n\) edges.
An \(n\)-club ensures all vertex pairs are connected by paths no longer than \(n\) edges.
The graph encoding approach of  qTKP and qMKP  can be applied to these structures. 
Additionally, 
the path length calculating and comparison for \(n\)-clan and \(n\)-club can be implemented using the degree count and degree comparison circuits of qTKP. 
Therefore, qTKP and qMKP demonstrate versatility and can be broadly applied to a range of problems related to cohesive structures in graphs. This adaptability enhances their utility across different graph-based computational challenges.

\section{\textcolor{XFcolor}{A Quantum Annealing Algorithm for MKP}}\label{sec:ann}
Quantum annealing represents another significant quantum computing model. The first commercially available device claiming to use quantum effects to accelerate optimization is the quantum annealer developed by D-Wave Systems. 
In contrast to the previously discussed gate-based quantum computing model, quantum annealing features a simpler and more comprehensible mathematical formulation. 
While gate-based models are more general, they may underutilize qubits for specific problems, a limitation that quantum annealing mitigates. 
To employ the D-Wave quantum annealer, a  combinatorial optimization problem should be formulated as a mathematical function of binary variables. The system then seeks assignments of these variables that minimize the function. This process uses quadratic unconstrained binary optimization (QUBO) encoding, which (1) allows only quadratic, or pairwise, interactions between variables, (2) without explicit constraints, (3)  uses only binary variables, and (4) encodes optimization problems. Physically, a QUBO problem can be viewed as an energy function, where the goal is to find the variable configuration corresponding to the minimum energy that represents an optimal solution.
In this section, we explore how to use the D-Wave quantum computer to solve the MKP by proposing a QUBO-based quantum annealing algorithm. 

\subsection{\textcolor{XFcolor}{Quadratic Unconstrained Binary Optimization}}
The initial step of applying a quantum annealing algorithm to solve the MKP is to formalize it as a QUBO. 
For clarity, we consider the problem of finding the maximum $k$-cplex on the complementary graph $\overline{G}$. We assign a binary variable $x_i=1$ if vertex $v_i$ appears in the solution, and $x_i=0$ otherwise. The objective is to maximize 
\(
\sum_{i=1}^n x_i
\), 
while ensuring that each selected vertex $v_i$ has no more than $k-1$ neighbors in the induced subgraph. Let 
\(
N(i) = \{j \neq i \mid (i,j) \in E\}
\) 
denote the set of indices of all neighbors of vertex $v_i$. The constraint can then be expressed as 
\begin{equation}\label{constraint1}
	\sum_{j \in N(i)} x_j \leq k-1, \quad \forall  i\in [1,n] \text{ with } x_i = 1
\end{equation}
QUBO formulation does not allow direct  constraints to the objective function. A standard approach is to convert constraints into a quadratic penalty term appended to the objective, so that any candidate solution violating the constraints is penalized heavily. The unconstrained objective function is thus given by 
\begin{equation}
	F = -\sum_{i=1}^n x_i + R \cdot p
\end{equation}
where $R$ is a constant parameter adjusting the penalty strength and $p$ is a quadratic penalty term. When $F$ attains its minimum, the corresponding $\{x_i\}$ represents the solution. 
Regarding the specific form of $p$, the  challenge is how to convert the inequality constraint into a quadratic penalty. Next, we discuss about the details. 

Since the requirement 
\(
\sum_{j \in N(i)} x_j \leq k-1
\) 
applies only when $x_i = 1$, we can uniformly write the inequality constraint for both $x_i = 1$ and $x_i = 0$ as 
\(
\sum_{j \in N(i)} x_j \leq k-1 + M(1-x_i),
\) 
where $M$ is a given large constant. It can be verified that for a sufficiently large $M$, the new inequality is equivalent to the original when $x_i = 1$, and imposes no restriction when $x_i = 0$. Thus, we transform the constraint into 
\begin{equation}\label{constraint2}
	\sum_{j \in N(i)} x_j \leq k-1 + M(1-x_i), \quad \forall\, i\in [1,n]
\end{equation}
This approach differs from Eq.~\ref{constraint1} as it no longer requires separate treatment for the cases $x_i = 1$ and $x_i = 0$, allowing for a unified  form.
Next, to convert this into an equality-form penalty, we introduce a slack variable $s_i \geq 0$ such that 
\begin{equation}\label{equlpe}
	\sum_{j \in N(i)} x_j + s_i - (k-1) - M(1-x_i) = 0
\end{equation}
By selecting 
\(
p_i = (\sum_{j \in N(i)} x_j + s_i - (k-1) - M(1-x_i))^2
\) 
as the penalty, 
$p_i = 0$  if and only if the original inequality constraint Eq.~\ref{constraint1} is satisfied; otherwise, $p_i > 0$.
Now the problem is that we cannot know the value of the slack variable $s_i$ in advance. In fact, $s_i$ is introduced as an unknown and is solved along with $\{x_i\}$. As long as the obtained $s_i$ is non-negative, constraint Eq.~\ref{constraint1} is satisfied. The approach is to represent $s_i$ with a binary form: 
\begin{equation}
	s_i = \sum_{r=0}^{L-1} 2^r s_{i,r}, \quad s_{i,r} \in \{0,1\}
\end{equation}
Here, $L$ is the number of bits sufficient to cover the maximum possible value of $s_i$. In this way, we restrict $s_i$ to non-negative integers and transform $s_i$ into a set of unknown binary variables $\{s_{i,r}\}$   similar to $\{x_i\}$, which are solved together with $\{x_i\}$.
Now we sum up all of the penalty terms $p = \sum_i p_i$ and get the final object function: 
\begin{equation}\label{objective}
	F = -\sum_{i=1}^n x_i + R \cdot \sum_{i=1}^n \Bigl(\sum_{j \in N(i)} x_j + s_i - (k-1) - M(1-x_i)\Bigr)^2
\end{equation}
Eq.~\ref{objective} is the QUBO form of the MKP we seek. We observe that $F$ contains only quadratic terms of binary variables $\{x_i, s_{i,r}\}$ and does not include explicit constraints. Thus, solving MKP is transformed into minimizing $F$.

\subsection{\textcolor{XFcolor}{Parameter Setting: $M, L$, and $R$}}

\subsubsection{Parameter $M$}
We observe that to make Eq.~\ref{constraint2} equivalent to Eq.~\ref{constraint1}, the chosen value of $M$ should be sufficiently large so that when $x_i = 0$, Eq.~\ref{constraint2} imposes no constraints on $\{x_i\}$. However, an excessively large $M$ enlarges the slack variable $s_i$, which in turn increases the number of binary digits $s_{i,r}$ needed to represent $s_i$, consuming additional variable resources. Therefore, our goal is to find a lower bound for $M$. We consider the smallest $M$ such that $\sum_{j \in N(i)} x_j \leq k-1 + M$ always holds. Let $d_{\overline G}(v_i)$ be the degree of vertex $v_i$ on the complementary graph $\overline{G}$. It can be verified that a lower bound for $M$ is 
\begin{equation}
	M = d_{\overline G}(v_i)-k+1
\end{equation}
which is our specific choice for $M$.

\subsubsection{Parameter $L$}
The size of integer $L$ is determined by the minimum number of binary bits $\{s_{i,r}\}$ needed to represent the integer slack variable $s_i$ that satisfies Equation \ref{equlpe}. In this context, we consider the maximum possible value of $s_i$, which splits into two cases: $x_i = 0$ and $1$. When $x_i = 0$, note that 
\(
s_i = (k-1) + M - \sum_{j \in N(i)} x_j = d_{\overline{G}}(v_i) - \sum_{j \in N(i)} x_j.
\) 
Here, $d_{\overline{G}}(v_i) \neq \sum_{j \in N(i)} x_j$ because $d_{\overline{G}}(v_i)$ denotes the global degree of vertex $v_i$ in $\overline{G}$, while $\sum_{j \in N(i)} x_j$ denotes the local degree of $v_i$ in the solution subgraph. Therefore, the upper bound for $s_i$ is 
\(
d_{\overline{G}}(v_i).
\) 
When $x_i = 1$, a similar analysis gives 
\(
s_i \leq k-1.
\) 
Combining these two cases, we have 
\begin{equation}
	s_i \leq  \max\{ d_{\overline{G}}(v_i),\; k-1\}
\end{equation}
Since $L$ is the number of binary bits required for $s_i$, we define 
\begin{equation}
	L = \lceil \log_2 \max\{ d_{\overline{G}}(v_i),\; k-1\} \rceil
\end{equation}
This is the specific choice for $L$.

\subsubsection{Parameter $R$}
Here, we discuss how to select  $R$ such that Equation  \ref{objective} is correct, meaning the global minimum of Equation  \ref{objective} corresponds to the solution of the MKP.
Assume the solution of the MKP is subgraph  $P$. Since  $P$ satisfies Equation  \ref{constraint1}, the corresponding value of Equation  \ref{objective} is  $F=-|P|$. For any subgraph  $P'$ with  $|P'|=|P|+t$, where  $t$ is a positive integer, we have 
\(
F'=-|P|-t+R\cdot p\geq -|P|-t+tR.
\) 
This inequality holds because, relative to  $P$, the subgraph  $P'$ contains at least  $t$ vertices that do not satisfy Equation  \ref{constraint1}, and each such vertex introduces a penalty of at least  $R$ to  $F'$. To ensure that  $F=-|P|$ is the global minimum, we require  $F' > F$, i.e., 
\(
-t+tR>0,
\) 
which implies 
\begin{equation}
	R>1
\end{equation}
$R$ plays a role   similar to the  regularization term in learning, of which the optimal value  can be determined experimentally.

\begin{algorithm}[tttt]
        \caption{Quantum Annealing for MKP   (qaMKP)}
        \begin{algorithmic}[1] 
            \Require Graph $G(V,E)$, $k$, $R$;
            \Ensure A maximum $k$-plex on graph $G$;
            \State Use a quantum annealer to find the minimum of  Eq.~\ref{objective}; 
            \State Output the maximum $k$-plex. 

        \end{algorithmic}
    \label{algo:qaMKP} 
   
    \end{algorithm}

\subsection{\textcolor{XFcolor}{Our Algorithm: qaMKP}}
Now we can summarize our quantum annealing  algorithm qaMKP   as  Algorithm~\ref{algo:qaMKP}. 
Similar to existing   annealing-based algorithms~\cite{10.14778/2947618.2947621,schonberger2022applicability}, the key  purpose   is to practically improve efficiency; therefore, its time complexity is equivalent to that of classical algorithms and is not analyzed separately.
We observe that the QUBO formulation of the MKP requires two groups of binary variables: vertex variables $\{x_i\}$ and slack variables $\{s_{i,r}\}$. When solving the QUBO with a quantum annealer, these binary variables are mapped to qubits. Ideally, each binary bit corresponds directly to one qubit; however, due to connectivity of D-Wave qubits, multiple qubits may be required to encode a single binary bit. As with gate-based quantum computers, qubit resources are scarce. Therefore, the number of binary variables in a QUBO problem is an important measure of algorithm performance. For Eq.~\ref{objective}, representing $n$ vertices requires $n$ bits. For each vertex, $L$ slack variables are needed. Thus, the total number of binary variables required by the algorithm is
\begin{equation*}
	n + nL =  n\Big(1+\lceil \log_2 \max\{ d_{\overline{G}}(v_i),\; k-1\} \rceil\Big) = O(n\log n)
\end{equation*}
Quantum annealing algorithms are progressive approximation algorithms with runtime provided as an input parameter and a solution that is progressively refined. The solution quality is measured by the cost   value. Note that for a   cost function with slack variables, the minimum cost may need not be reached to obtain the optimal solution; 
the quantum annealer may find the optimal solution without optimally configuring the slack variables. This is acceptable because slack variables are auxiliary and their optimal configuration is unnecessary.



\section{Experimental Studies}\label{sec:4}
In this section, we perform proof-of-principle experiments using state-of-the-art IBM quantum simulators and D-Wave adiabatic quantum computers. We conduct experiments on qTKP/qMKP and qaMKP separately. Regarding qTKP/qMKP, we investigate the following research questions: 
(1) In the iterative process of the oracle in Figure~\ref{qcfull} for qTKP, can the amplitude of the solution converge rapidly? 
(2) How does the performance of qMKP compare with classical algorithms, and what is the relationship between efficiency and graph size? 
(3) How quickly can qMKP, as a progressive algorithm, find the first feasible solution that is at least half the size of the optimal solution? 
(4) During the operation of the oracle, what patterns emerge in the proportion of runtime occupied by each of its three components? 
(5) How do the answers to the above four questions change when varying the value of \( k \)?
Regarding qaMKP, we investigate the following research questions: 
(6) Since adiabatic quantum computers perform multiple times of annealing, how should the time for each annealing be set in practice? 
(7) What is a suitable value of \( R \) (penalty strength) in the qaMKP algorithm for best performance? 
(8) How does the performance of qaMKP compare with classical algorithms, and what is the relationship between efficiency and graph size? 
(9) How does the performance of qaMKP change when varying the value of \( k \)? 
(10) D-Wave devices require multiple qubits to represent a binary variable. The average number of qubits  used for each bit  is called the chain strength. How does the chain strength change with increasing graph size?
All experiments were conducted using Python 3.8. 
qTKP and qMKP were tested on IBM simulators MPS with Qiskit. 
qaMKP was tested on D-Wave Advantage System 5.4 (QPU solver) and Hybrid Binary Quadratic Model Version 2 (hybrid solver). 
All other classical algorithms and quantum algorithms with classical implementations were tested on  a MacBook Pro with 32GB memory and an Intel Core i7 2.6 GHz CPU~\footnote{Source code is available at  https://github.com/XFLi25/qMKP}.

\begin{table}
\renewcommand\arraystretch{1.2}
\centering
\caption{Dataset sizes of existing quantum  database works}
\vspace{-6pt}
\begin{tabular}{p{76pt}p{113pt}p{2pt}p{4pt}}
\toprule
Problem  & Time complexity \& Work &$n$ & $m$      \\
\midrule
Maximum clique & $O^*(2^{\frac{n}{2}})$~\cite{chang2018quantum}&  2  &  4  \\
$k$-clique  & $O^*(2^{\frac{n}{2}})$~\cite{metwalli2020finding}&  4 &   4   \\
Maximum $k$-plex  &$O^*(2^{\frac{n}{2}})$ [qMKP] &  10 &   23   \\
Maximum $k$-plex  &$\quad\backslash \quad\quad$ [qaMKP] &  30 &   300   \\
\bottomrule
\end{tabular}
\label{datasize}
\vspace{-15pt}
\end{table}

\begin{figure*}[t]
\vspace{-26pt}
	\subfloat[Before iteration \label{expa}]{\includegraphics[width=4.6cm]{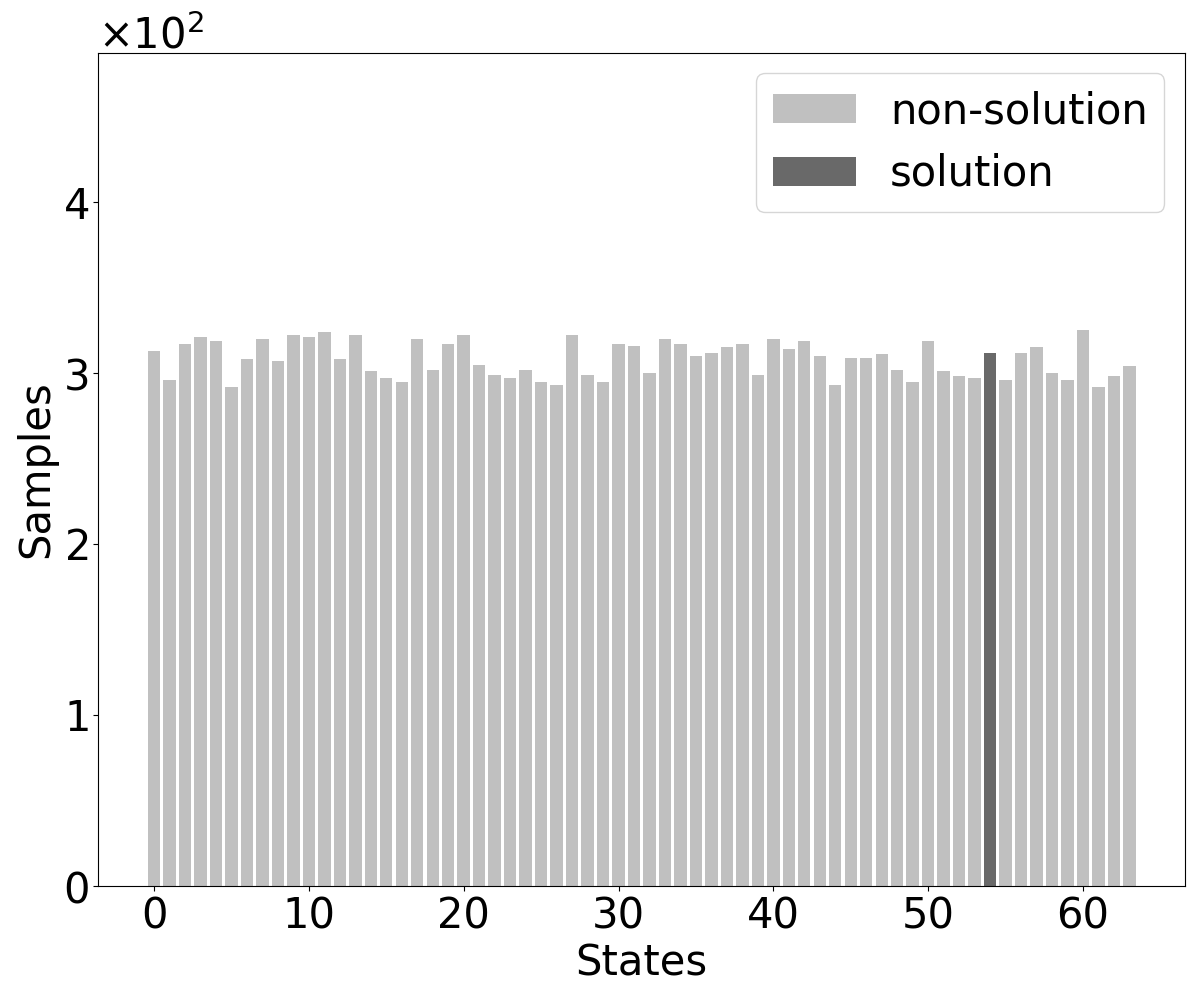}}
	\subfloat[Iteration 1 \label{expb}]{\includegraphics[width=4.6cm]{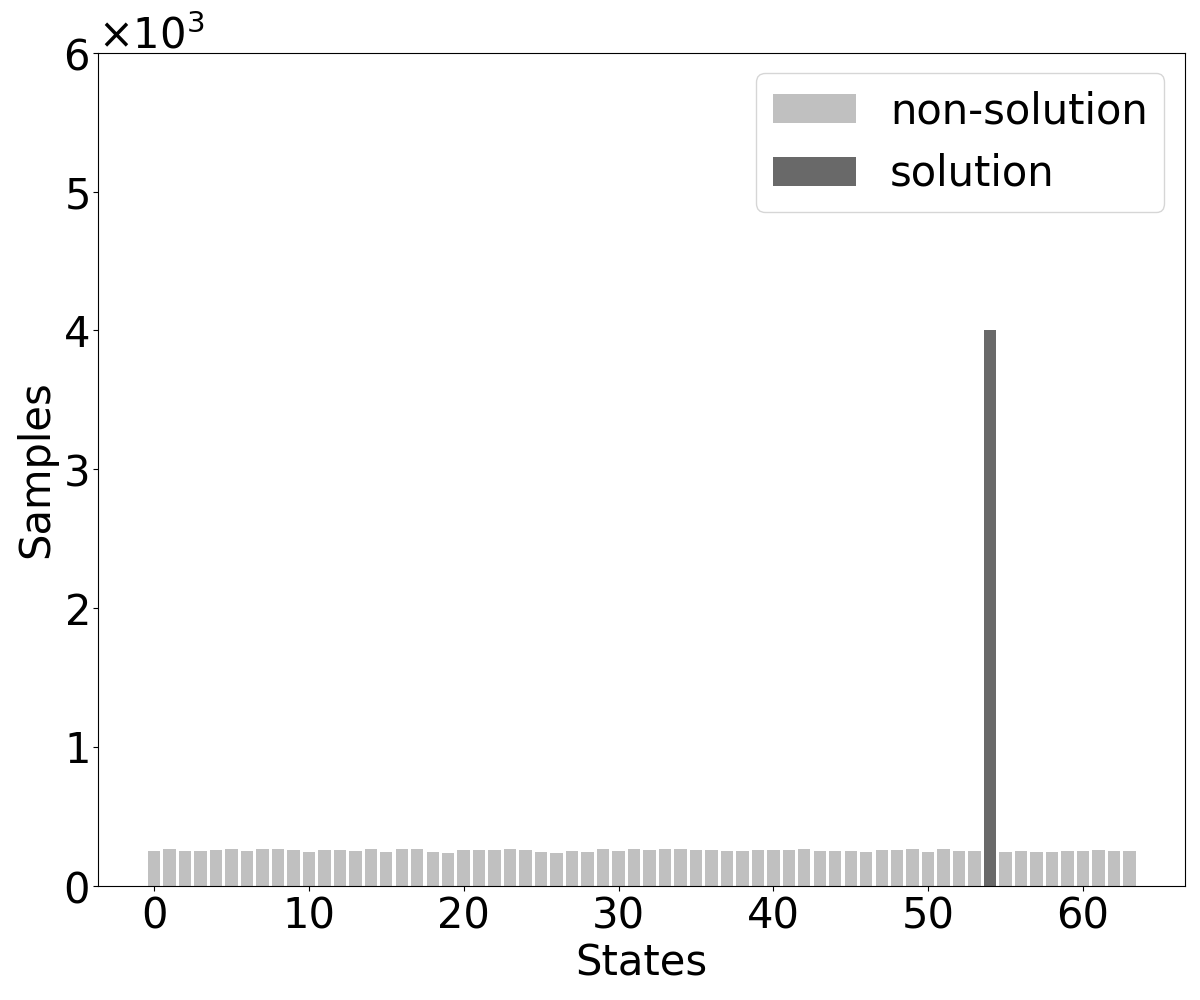}}
	\subfloat[Iteration 3 \label{expc}]{\includegraphics[width=4.6cm]{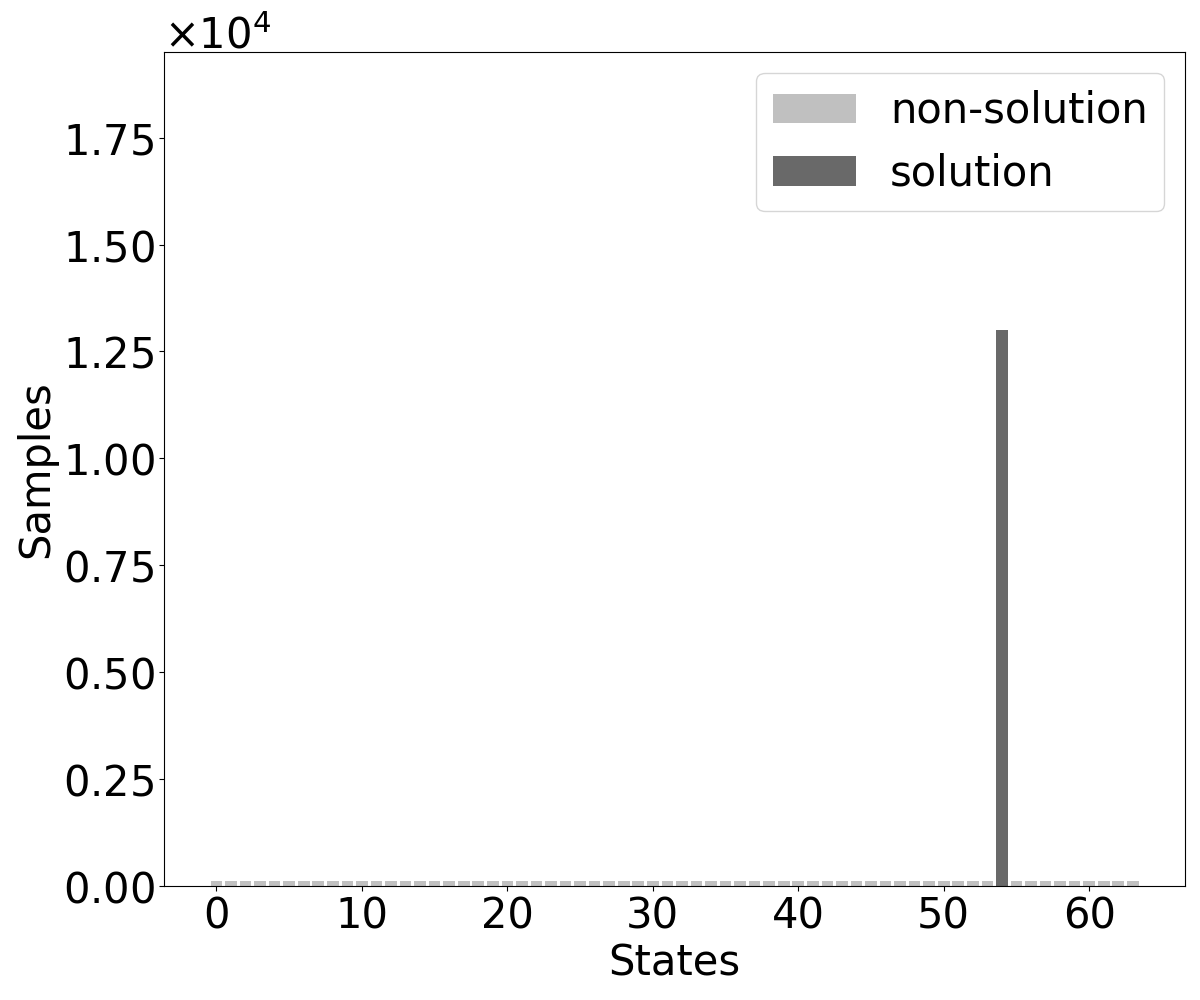}}
\subfloat[Iteration 6 (final)  \label{expd}]{\includegraphics[width=4.6cm]{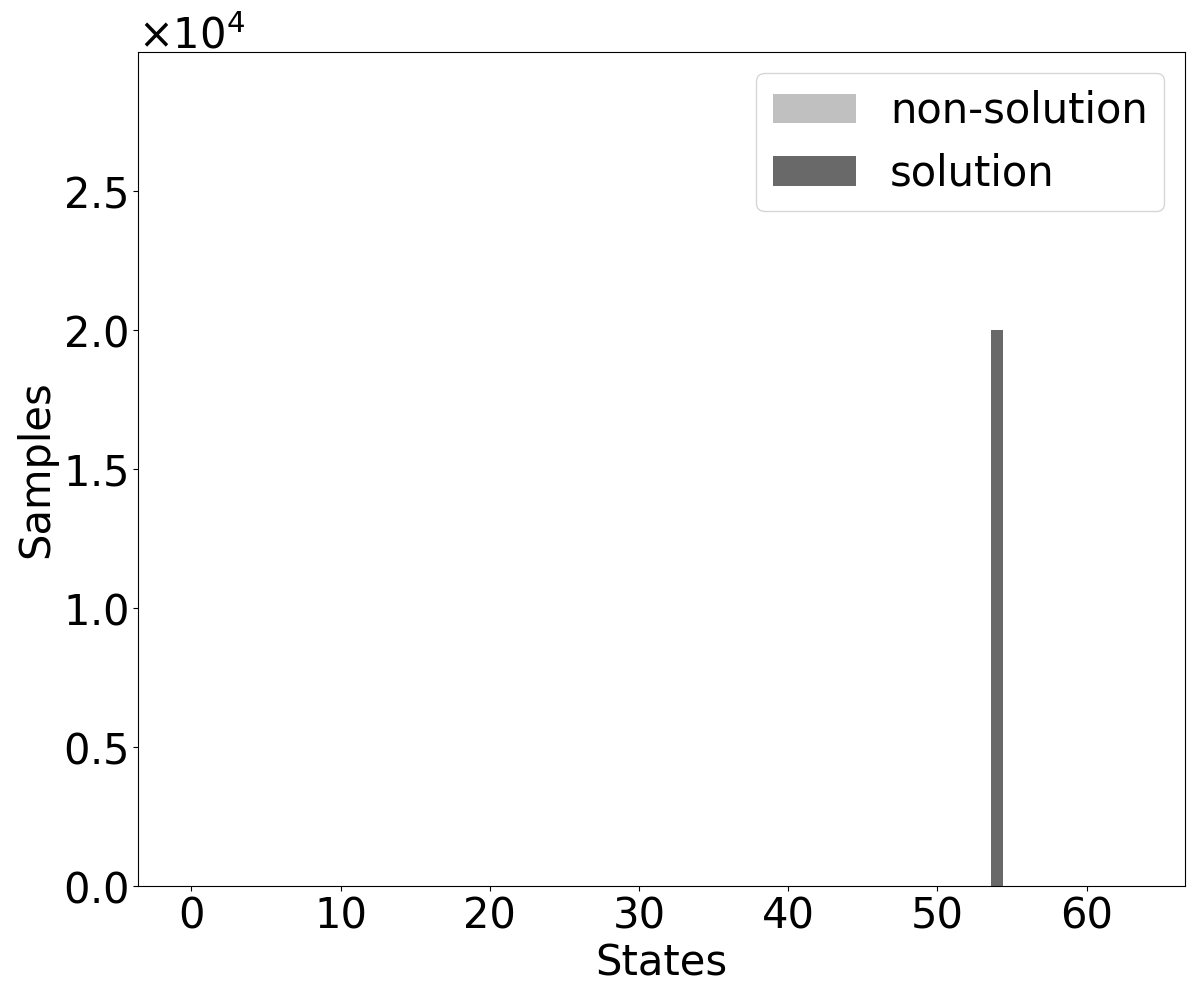}}	

\vspace{-3pt}

	\captionsetup{justification=centering}
	\caption{Subgraph amplitude distribution in the running process of qTKP}\label{fig:exp}
\vspace{-16pt}
\end{figure*}

\subsection{Error Probability Convergence of qTKP}
In quantum computing, indeterminacy leads to an inherent non-zero error probability where the final state may collapse incorrectly upon measurement. This fundamental aspect cannot be eliminated theoretically. The error probability is bounded by \(\pi^2/(4I)^2\), where \(I\) is the number of iterations~\cite{nielsen_chuang_2010}. For a small \(I\), running the qTKP algorithm \(c\) times further reduces the error probability to \(\pi^2/(4I)^{2c}\). Thus, the error rate decreases rapidly, which supports the practical viability of our algorithms for yielding accurate solutions. 
To examine how the error rate converges in practice, we executed qTKP with 20K shots on the graph shown in Figure~\ref{graph}. We measured the intermediate states during algorithm execution and reported the frequency distribution across 64 base states, ranging from \(\ket{000000}\) to \(\ket{111111}\).

We conducted a case study on the intermediate states of the qTKP algorithm running on the graph in Figure~\ref{graph} to examine the convergence of the error probability. This graph is larger than those typically used in existing quantum graph database works~\cite{chang2018quantum,metwalli2020finding}, as detailed in Table~\ref{datasize}.
The experimental results in Figure~\ref{fig:exp} show the frequency distribution of 64 subgraphs over 20,000 runs. The horizontal axis represents the 64 bases corresponding to subgraphs, with the solution marked in dark gray. The vertical axis shows how often the intermediate state collapsed to each subgraph, effectively illustrating a probability distribution chart.
Initially, as shown in Figure~\ref{expa}, the amplitudes of the 64 subgraphs are uniform, forming an equal superposition state. After one oracle iteration (Figure~\ref{expb}), the solution amplitude becomes significantly higher than the non-solutions, with a correct result probability of 20.5\% (error probability of 79.5\%). Figure~\ref{expc} shows the state after three iterations, where non-solution amplitudes are nearly zero and the error probability is 35.3\%. Figure~\ref{expd} presents the final state after six iterations, reducing the error probability to 0.075\%.
From Figure~\ref{fig:exp}, the error probability converges rapidly, becoming negligibly small after six iterations. The convergence rate is approximately inversely proportional to the square of the number of iterations. For larger datasets requiring more iterations, the error probability quickly falls. These findings support that although qMKP has a random component, it operates effectively as an exact algorithm in practice.

\begin{table}
\renewcommand\arraystretch{1.2}
\centering
\caption{Performance of the qMKP   with \( k=2 \) on datasets varying sizes}
\vspace{-6pt}
\begin{tabular}{l l l l l }
\toprule
Dataset  & $G_{7,8}$ & $G_{8,10}$ &$G_{9,15}$  &$G_{10,23}$  \\
\midrule
Maximum $k$-plex size&4&4&5&6\\
BS (\SI{}{\micro\second})&327.4&335.6&343.5&353.7\\
qMKP (\SI{}{\micro\second})&126.4&132.8&138.6&130.3\\
First-result time (\SI{}{\micro\second})&34.6&37.5&39.6&35.7\\
First-result size &3&3&3&4\\
Error probability & $<10^{-4}$& $<10^{-5}$& $<10^{-6}$& $<10^{-6}$\\

\bottomrule
\end{tabular}
\label{compsize}
\vspace{-15pt}
\end{table}

\subsection{Comparison of qMKP with state-of-the-art on datasets of varying sizes}
Given that the quantum simulator available to us supports up to 100 qubits,
although our algorithm outperforms existing methods in complexity, current QPUs limit testing on large datasets. We thus compare our algorithm with the BS algorithm~\cite{xiao2017fast} on small   datasets. We select the BS algorithm as baseline due to its non-trivial time complexity. 
To test qMKP on slightly larger datasets, we integrated the core-truss co-pruning technique~\cite{chang2022efficient}, making the datasets suitable for current simulators after graph reduction. For broader applicability, we evaluated four synthetic datasets with vertex counts from 7 to 10, significantly larger than those in previous studies~\cite{chang2018quantum,metwalli2020finding}. We denote a dataset as \( G_{i,j} \), where \( i \) is the vertex count and \( j \) the edge count. Results are shown in Table~\ref{compsize} with \( k \) fixed at 2. Reported running times average over 20K executions.


Table~\ref{compsize} shows that qMKP is 2.5-2.7 times faster than BS on all datasets. qMKP performs slightly better on denser graphs (e.g., \(G_{10,23}\)) than on sparser ones (e.g., \(G_{9,15}\)), since it processes the complementary graph more effectively. As a progressive algorithm, qMKP provides near-optimal intermediate results during binary search and outputs a feasible solution at least half the size of optimal in under 30\% of total time. 
Furthermore, the error probability decreases exponentially as vertex count increases, aligning with the theoretical limit of \(\pi^2/(4I)^2\). For the 10-vertex dataset, the error probability reduces to a negligible \(10^{-6}\).



\subsection{Performance of qMKP with varying values of $k$}

In practice, \( k \) is usually small (at most 5)~\cite{gao2018exact,xiao2017fast,zhou2021improving}. We tested qMKP for \( k = 2, 3, 4, 5 \) on a selected dataset, with results shown in Table~\ref{compk}. As \( k \) increases, qMKP running time rises slightly, about a 7\% increase from \( k=2 \) to \( k=5 \). This is because a higher \( k \) increases resource consumption only in the degree comparison part of the oracle, which is a minor portion of its overall workload. qMKP maintains a speedup of approximately 2.6-2.7 times compared to BS regardless of \( k \). Both the first-result time and size are nearly independent of \( k \). The error probability remains unchanged with increasing \( k \) since the number of iterations does not depend on \( k \).

\begin{table}
\vspace{-15pt}
\renewcommand\arraystretch{1.2}
\centering
\caption{Performance of the qMKP   on the \( G_{10,37} \) dataset for \( k=2, 3, 4, 5 \)}
\vspace{-5pt}
\begin{tabular}{l l l l l }
\toprule
$k$  & $2$ & $3$ &$4$  &$5$  \\
\midrule
Maximum $k$-plex size&6&6&6&7\\
BS (\SI{}{\micro\second})&353.7&357.3&363.4&374.8\\
qMKP (\SI{}{\micro\second})&130.3&131.4&137.5&139.6\\
First-result time (\SI{}{\micro\second})&35.7&37.5&37.3&36.7\\
First-result size &4&4&4&4\\
Error probability & $<10^{-6}$& $<10^{-6}$& $<10^{-6}$& $<10^{-6}$\\

\bottomrule
\end{tabular}
\label{compk}
\vspace{-5pt}
\end{table}

\begin{table}
\renewcommand\arraystretch{1.2}
\centering
\caption{Proportional share of three oracle components in the running time of the qMKP   across various datasets}
\vspace{-5pt}
\begin{tabular}{l l l l l }
\toprule
Dataset  & $G_{7,8}$ & $G_{8,10}$ &$G_{9,15}$  &$G_{10,23}$  \\
\midrule
Degree count (\%)&77.5&82.6&88.5&88.6\\
Degree comparison (\%)&10.9&8.3&5.3&5.4\\
Size determination (\%)&11.6&9.1&6.2&6.0\\

\bottomrule
\end{tabular}
\label{per}
\vspace{-15pt}
\end{table}

\subsection{Runtime allocation of different  oracle parts of qMKP}

In qMKP, oracle iteration is the core, and most of the runtime is spent in the oracle. We experimentally measured the runtime proportions of the three oracle components (Table~\ref{per}) to guide future optimization. The degree count part dominates the oracle operation, accounting for 77.5\% to 93.5\% of the time, and its share grows significantly with the number of vertices. The other two components consume similar amounts of time. The high time share of degree count is due to its \( O(n^3 \log n) \) complexity compared to \( O(n \log n) \) for the others, and this share increases as \( n \) grows.
This explains why with $k$ increasing from 2 to 5, qMKP experiences only a marginal increase in running time, as $k$ only primarily impacts the degree comparison component. 



\begin{table}
\vspace{-15pt}
\renewcommand\arraystretch{1.2} 
\centering
\caption{\textcolor{XFcolor}{Comparison between different annealing time $\Delta t$ with different datasets ($k = 3, R = 2$)}}
\vspace{-5pt}
\begin{tabular}{l l l l l l l}
\toprule
Dataset    & $1\,\mu$s & $10\,\mu$s & $20\,\mu$s & $40\,\mu$s & $100\,\mu$s & $200\,\mu$s \\
\midrule
$D_{10,40}$  & \textbf{3}    & 8    & 11   & 18   & 17   & 18   \\
$D_{15,70}$  & \textbf{36}   & 53   & 47   & 45   & 91   & 114  \\
$D_{20,100}$ & \textbf{398}  & 526  & 756  & 974  & 997  & 553  \\
$D_{30,300}$ & \textbf{1413} & 1514 & 1796 & 1937  & 2035 & 1179  \\
\bottomrule
\end{tabular}
\label{tab:annealing_time}
\vspace{-5pt}
\end{table}


\begin{table}
\renewcommand\arraystretch{1.2}
\centering
\caption{\textcolor{XFcolor}{Comparison between different $R$'s on $D_{10,40}$ ($k = 3, \Delta t = 1\,\mu$s)}}
\vspace{-5pt}
\begin{tabular}{l l l l l l l l}
\toprule
   $R$    & $1\,\mu$s       &  $5\,\mu$s        & $10\,\mu$s         & $50\,\mu$s        & $100\,\mu$s       & $500\,\mu$s       & $1000\,\mu$s \\ 
\midrule
$1.1$ & 6.4      & 11.8    & 0.9   & \textbf{-3.4}   & \textbf{-3.4}      & \textbf{-5.6}   & \textbf{-6.7} \\
$2$   & \textbf{0}        & 24      & \textbf{-4}         & \textbf{2}      & \textbf{0}      & \textbf{-6}     & \textbf{3}   \\
$4$   & 36       & 55      & 23         & 15      & \textbf{2}         & \textbf{6}      & \textbf{2}    \\
$8$   & 208      & {102}    & 79         & 23      & 23        & \textbf{22}    & \textbf{6}    \\
\bottomrule
\end{tabular}
\label{tab:r}
\vspace{-15pt}
\end{table}


\subsection{\textcolor{XFcolor}{Annealing time  $\Delta t$ of qaMKP}}
In quantum annealing, runtime is a parameter. What we test is whether the objective cost can decrease as the runtime parameter increases, thus obtaining the solution corresponding to the global minimum of the cost. The overall QPU runtime depends on the annealing time per shot \( \Delta t \) and the number of shots \( s \) performed by the quantum computer, so the total runtime is \( t = \Delta t \cdot s \). 
Then the   question is: to observe the  performance of qaMKP on the D-Wave Advantage System 5.4 with a given runtime \( t \), should we select a larger \( \Delta t \) with a smaller \( s \), or vice versa? Existing work~\cite{10.14778/2947618.2947621,trummer2024leveraging,schonberger2023ready} generally fixes \( \Delta t \) to a constant value, typically between $1\,\mu\text{s}$ and $129\,\mu\text{s}$. 
In this work, we test qaMKP on datasets of different sizes with runtime \( t = \Delta t \cdot s = 1000\,\mu\text{s} \) under varying \( \Delta t \). 
We selected four synthetic datasets $D_{n,m}$: \( D_{10,40}, D_{15,70}, D_{20,100}, D_{30,300} \) and varied \( \Delta t \) between $1\,\mu\text{s}$ and $200\,\mu\text{s}$, using default parameters \( k = 3 \) and \( R = 2 \). We recorded the objective cost obtained upon completion of qaMKP and present the results in Table~\ref{tab:annealing_time}. 
We observed that for all datasets, the cost generally increases as \( \Delta t \) increases. Some exceptions occur because a larger \( \Delta t \) reduces sample times \( s \), leading to more fluctuation in cost uncertainty. The minimum/optimal cost consistently occurred at \( \Delta t = 1\,\mu\text{s} \). This indicates that for datasets of this scale, an annealing time of $1\,\mu\text{s}$ is sufficient to obtain good results, and runtime adjustments should focus on varying the sample times \( s \). In subsequent experiments, we set the default \( \Delta t = 1\,\mu\text{s} \).

\begin{figure}[t]
\vspace{-20pt}

	\centering
	{\includegraphics[width=8cm]{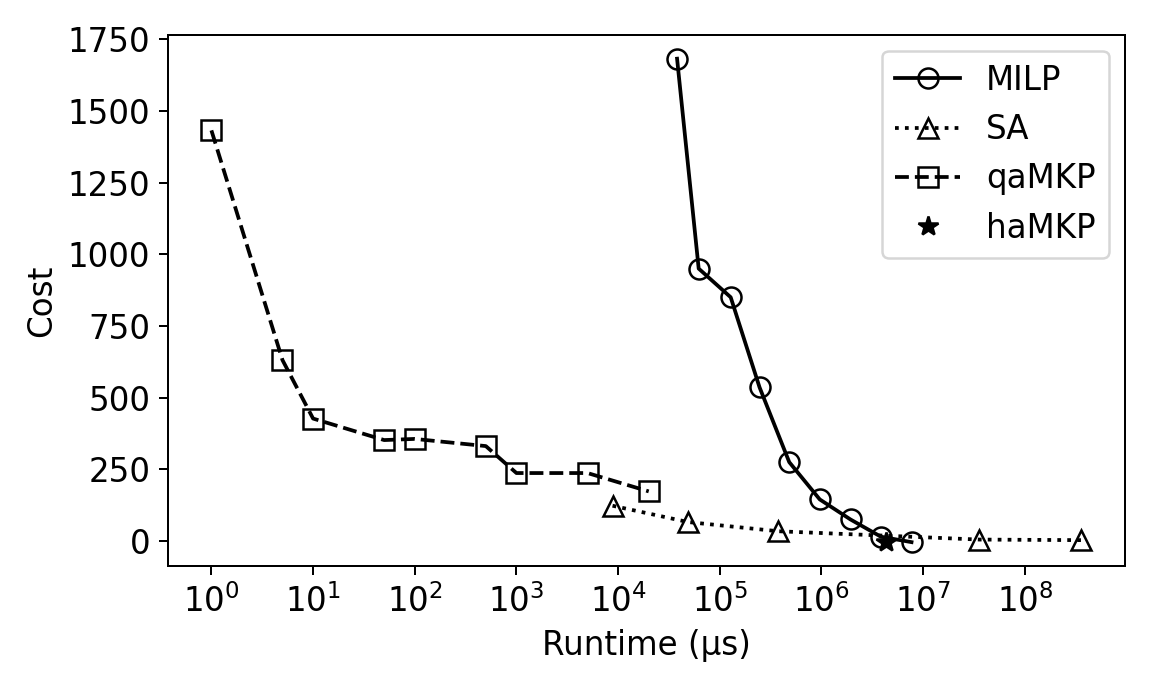}}

	\vspace{-10pt}
	\captionsetup{justification=centering}
	\caption{\textcolor{XFcolor}{Objective cost as a function of runtime for different algorithms on dataset \(D_{20,100}\), default settings \(k = 3\), \(R = 2\), \(\Delta t = 1\,\mu\text{s}\)}\label{runtime1}}
	\vspace{-15pt}
\end{figure}

\subsection{\textcolor{XFcolor}{Penalty parameter $R$ of qaMKP}}
We have proven that to guarantee the correctness of qaMKP, \( R \) must satisfy \( R > 1 \). Here, \( R \) represents the penalty strength applied to the penalty term in qaMKP. A larger \( R \) enforces the \( k \)-plex constraint more strictly, while a smaller \( R \) places greater emphasis on optimizing the size of the solution. We experimentally evaluate the performance of qaMKP under different \( R \) values. We increase the runtime from \( 1\,\mu\text{s} \) to \( 1000\,\mu\text{s} \) and observe which of four \( R \) values (1.1, 2, 4, 8) finds the optimal solution first. We select dataset \( D_{10,40} \) with default settings \( k = 3 \) and \( \Delta t = 1\,\mu\text{s} \), increasing the total runtime by raising the number of shots \( s \). The results are shown in Table~\ref{tab:r}. We record the objective cost for each experiment and use boldface to indicate when the optimal solution is found. Note that the optimal solution does not necessarily correspond to the minimum objective cost because slack variables may not be optimized to yield a penalty term of zero, but this does not affect the experimental outcome. We observe that \( R = 2 \) performs best; qaMKP finds the optimal solution at \( 1\,\mu\text{s} \), whereas for \( R = 1.1, 4, 8 \), the optimal solution is found at \( 50\,\mu\text{s} \), \( 100\,\mu\text{s} \), and \( 500\,\mu\text{s} \), respectively. This aligns with our intuition that although \( R \) must be greater than 1, it should not deviate far from 1. In Eq.~\ref{objective}, the quadratic penalty term multiplied by \( R \) can easily become much larger than the graph size when any vertex does not satisfy the constraint. In other words, the penalty term is inherently severe. In subsequent experiments, we set the default \( R = 2 \).

\subsection{\textcolor{XFcolor}{Comparison of qaMKP with classical algorithms}}

We compare four algorithms or implementations for MKP: qaMKP on D-Wave Advantage System 5.4 (denoted as qaMKP), qaMKP on D-Wave Hybrid Binary Quadratic Model Version 2 (denoted as haMKP), a Simulated Annealing algorithm on a local device (classical algorithm, denoted as SA), and Mixed-Integer Linear Programming using the state-of-the-art Gurobi Optimizer on a local device (classical algorithm, denoted as MILP). 
The first three algorithms optimize the same objective function Eq.~\ref{objective}. For MILP, each pair of binary variables \(X_u \cdot X_v\) in Eq.~\ref{objective} 
is linearized by introducing a new variable \(y_{u,v}\) in Gurobi subject to the constraints 
\[
y_{u,v} \leq X_u,\; y_{u,v} \leq X_v,\; y_{u,v} \geq X_u + X_v - 1,\; y_{u,v} \geq 0.
\] 
Diagonal terms \(X_u^2 = X_u\) remain unchanged. Based on this linearization, we construct an MILP objective 
\begin{equation}\label{milp}
	F = \text{offset} + \sum_{(u,v)} Q_{u,v} \cdot Z_{u,v}
\end{equation}
where \(Z_{u,v} = X_u\) if \(u = v\), or \(y_{u,v}\) if \(u < v\). The Gurobi Optimizer is then used to optimize Eq.~\ref{milp}.

\begin{figure}[t]
\vspace{-20pt}
	\centering
	{\includegraphics[width=8cm]{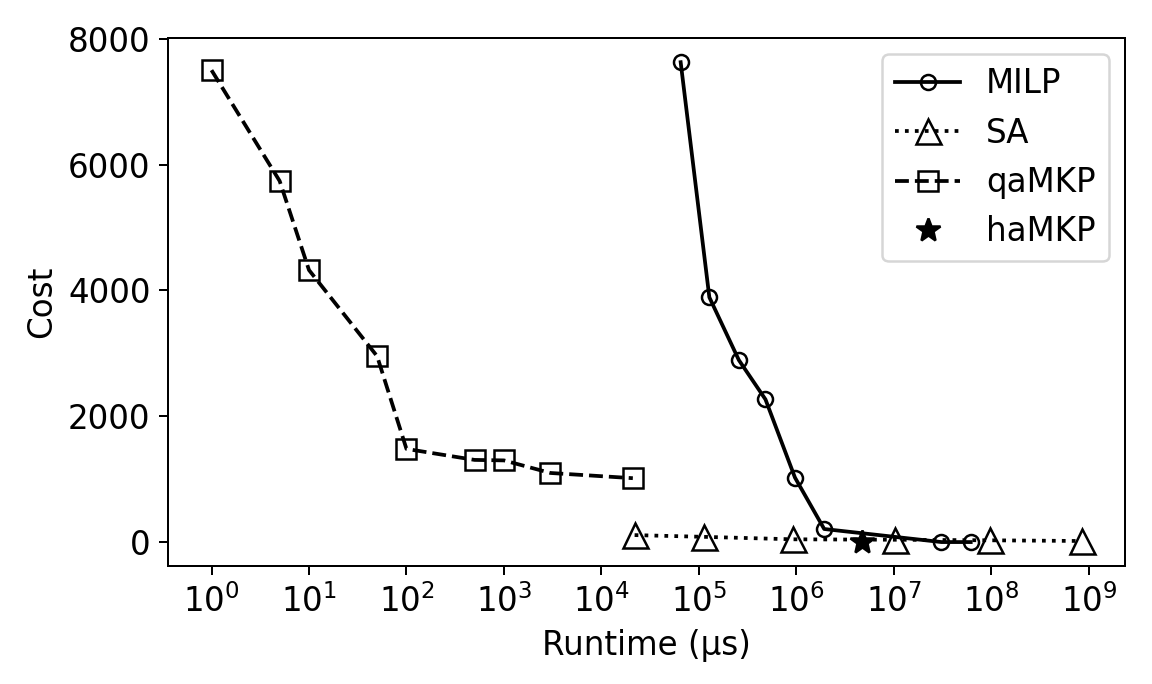}}

	\vspace{-10pt}
	\captionsetup{justification=centering}
	\caption{\textcolor{XFcolor}{Objective cost as a function of runtime for different algorithms on dataset \(D_{30,300}\), default settings \(k = 3\), \(R = 2\), \(\Delta t = 1\,\mu\text{s}\)}\label{runtime2}}
	\vspace{-15pt}
\end{figure}

The four algorithms employ different methods to control runtime. For qaMKP, the annealing time \( \Delta t \) is fixed while the number of shots \( s \) is adjusted. For haMKP, a minimum runtime of 3 seconds is directly specified. For SA, similar to qaMKP, the number of sweeps (analogous to annealing time) and the number of shots \( s \) must be set; we fix the number of sweeps to 2 and vary \( s \) to control runtime. For MILP, a runtime limit is directly set. We test the algorithms on two datasets \( D_{20,100} \) and \( D_{30,300} \), and the curves of objective cost versus runtime are shown in Fig.~\ref{runtime1} and Fig.~\ref{runtime2}. The horizontal axis represents runtime (in microseconds, log scale) and the vertical axis represents cost. Because haMKP requires at least 3 seconds of runtime and almost always finds a solution within this period, the figures display only a single data point marked with a star  for haMKP. 
We set default   \( k = 3 \) and \( R = 2 \).

In Fig.~\ref{runtime1} for \(D_{20,100}\), qaMKP reduces the cost from 1432 to 173 within \(1\,\mu\text{s}\) to \(10^4\,\mu\text{s}\), with the decrease trend gradually slowing. (Due to a maximum call time per QPU on the D-Wave Advantage System 5.4, further runtime increases are not allowed.) MILP reduces the cost from 1682 to $-4$ between \(10^4\,\mu\text{s}\) and \(10^7\,\mu\text{s}\), with a similarly slowing trend. SA decreases the cost from $123$ to $3$ between \(10^4\,\mu\text{s}\) and \(10^9\,\mu\text{s}\), with a consistently slow descent. haMKP finds a solution with cost $-4$ at approximately \(10^6\,\mu\text{s}\).
In terms of finding the optimal solution, MILP and haMKP perform best, as both can find the optimal solution around \(10^6\,\mu\text{s}\). 
This result is expected as Gurobi is a state-of-the-art MILP solver and the 
D-Wave hybrid service is optimized for QUBO problems, 
combining quantum computing with classical supercomputing advantages.
By contrast, qaMKP can quickly find a good sub-optimal solution in less than \(10^4\,\mu\text{s}\). At \(10^4\,\mu\text{s}\), qaMKP significantly outperforms Gurobi (MILP) and is roughly comparable to SA. Notably, to find a solution with cost below 500, qaMKP is about four orders of magnitude faster than Gurobi (MILP). This demonstrates that for a range of MKP problems, the quantum annealer could outperform the classical approach with a speedup of more than \(10^4\), aligning with observations in existing work~\cite{10.14778/2947618.2947621}.

The results shown in Fig.~\ref{runtime2} for \(D_{20,100}\) are very similar to those in Fig.~\ref{runtime1}, with the main difference being a less significant convergence of qaMKP. At \(10^4\,\mu\text{s}\), qaMKP converges to 1010, while SA reaches 104.
This weakening occurs because as the graph size \(n\) increases, the number of qubits required to represent a binary variable in Eq.~\ref{objective} gradually increases, which can affect the performance of the quantum annealer. This aspect will be discussed in detail later.

\begin{table}[t]
\vspace{-15pt}
\renewcommand\arraystretch{1.2}
\centering
\caption{\textcolor{XFcolor}{Objective cost as a function of runtime for qaMKP under different \(k\) values on  \(D_{20,100}\), default \(R = 2\), \(\Delta t = 1\,\mu\text{s}\)}}
\vspace{-5pt}
\begin{tabular}{
  p{0.15cm}  
  p{0.4cm}  
  p{0.4cm}  
  p{0.45cm}  
  p{0.4cm} 
  p{0.6cm}  
  p{0.6cm}  
  p{0.7cm}  
  p{0.8cm}  
}
\toprule
$k$      &  $1\,\mu$s       & $5\,\mu$s      & $10\,\mu$s       & $50\,\mu$s       & $100\,\mu$s      & $500\,\mu$s      & $1000\,\mu$s     & $4000\,\mu$s     \\
\midrule
$2$  & 1310   & 1013   & 927   & 633    & 373    & 366    & 275    & 234    \\
$3$  & 1432  & 631    & 427      & 352      & 356      & 331      & 237      & 246      \\
$4$  & 490      & 490      & 388      & 322      & 313      & 256      & 196      & 134      \\
$5$  & 820      & 493      & 412      & 339      & 325      & 222      & 200      & 93       \\
\bottomrule
\end{tabular}
\label{tab:k}
\vspace{-12pt}
\end{table}

\subsection{\textcolor{XFcolor}{Performance of qaMKP with varying values of $k$}}
We evaluate how qaMKP performance varies with different \(k\) values. Using dataset \(D_{20,100}\), we set \(k = 2, 3, 4, 5\) and \(\Delta t = 1\,\mu\text{s}\), observing changes in cost as runtime increases. The results are shown in Table~\ref{tab:k}. For each \(k\), the cost clearly decreases as runtime increases. Comparing different \(k\) values, we do not observe a distinct pattern, similar to  qMKP, where the value of \(k\) does not significantly affect performance. This is because qaMKP essentially searches for solutions in a space of size \(O(2^n)\), which is independent of \(k\).

\subsection{\textcolor{XFcolor}{Chain strength of qaMKP on D-Wave}}
\begin{figure}[t]
\vspace{-20pt}

	\centering
	{\includegraphics[width=8cm]{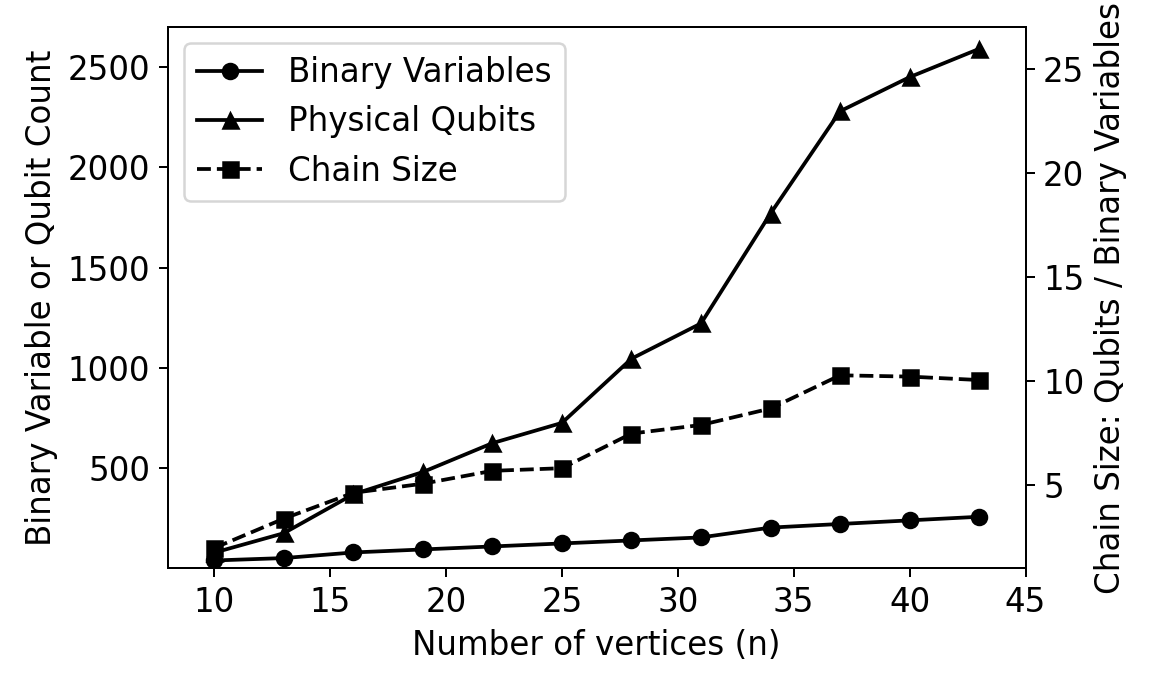}}

	\vspace{-10pt}
	\captionsetup{justification=centering}
	\caption{\textcolor{XFcolor}{Variable counts and chain size  vs graph size $n$}}\label{chain}
	\vspace{-15pt}
\end{figure}

In a physical embedding, each binary variable is mapped to a connected group of qubits. First, qubits representing the same variable must be connected to ensure they ultimately share the same value. Second, if two variables appear together in a quadratic term of the QUBO formula, at least one qubit from the first variable must connect to one from the second. The set of physical qubits representing a logical variable is called a chain. The embedding problem is NP-hard; therefore, we adopt a heuristic approach~\cite{cai2014practical}. Next, we report how binary variable count, qubit count, and average chain size change with increasing graph size from $n=10$ to $n=43$.  We set default $k = 3$, $R = 2$. Results are shown in Fig.~\ref{chain}. The left and right vertical axes represent variable count and average chain size, respectively. The number of binary variables required for qaMKP increases slowly from 40 to 258, strictly conforming to an $O(n\log n)$ growth trend. The number of physical qubits increases relatively quickly from 79 to 2591. The average chain size rises from 2 to around 10. This trend is intuitive: as $n$ increases, the number of quadratic terms that require pairwise connections increases, which on average necessitates more qubits to represent the same binary variable. For a quantum annealer, a larger chain size impedes further reduction of cost~\cite{liu2019solving}, which explains the gap between qaMKP and SA at runtime  $10^4\,\mu$s in Fig.~\ref{runtime2}. Despite limitations imposed by chain size, the $O(n\log n)$ complexity, independent of the number of edges $m$, still allows qaMKP to remain applicable to graphs much larger than those manageable by qMKP and unaffected by graph density.


\subsection{Summary}
We conducted tests of the graph-based qTKP/qMKP and the annealing-based qaMKP on the IBM quantum simulator and the D-Wave adiabatic quantum computer, respectively, and compared them with classical algorithms. 
For qTKP/qMKP, we observed that the amplitude of the solution converges rapidly and eventually becomes very close to 1, and it demonstrates comparable efficiency to the classical method. 
qaMKP is more suitable for larger and denser graph sizes relative to qTKP/qMKP. 
The objective cost converges rapidly in the early stage, and in some scenarios (moderate cost thresholds), qaMKP surpasses the classical MILP in efficiency by several orders of magnitude. Note that the criteria for fair QPU-CPU comparisons are a subject of scientific debate~\cite{mcgeoch2019principles}. Therefore, such comparisons are only used to demonstrate the potential of QPU advantages to the DB community.

\section{Related Works}\label{sec:5}

The literature related to our work can generally be categorized into the following three main groups:

\textbf{Exact Solutions for MKP.}
Exact search algorithms for MKP are tools used to find the optimal solutions by thoroughly checking through all options in a search space. 
Balasundaram et al.~\cite{balasundaram2011clique} formulated an integer programming model and a branch-and-cut algorithm for MKP.
McClosky and Hicks~\cite{mcclosky2012combinatorial} introduced a Branch-and-Bound (BnB) algorithm, modifying a maximum clique algorithm with a co-k-plex coloring for an upper bound.
Moser et al.~\cite{moser2012exact} developed an algorithm specifically targeting the maximum \( k \)-cplex.
Xiao et al.~\cite{xiao2017fast} presented a branch-and-search algorithm leveraging MKP's structural properties to streamline the search, achieving a time complexity of \( O(c^n n^{O(1)}) \) where \( 1.94 \leq c < 2 \) for \( k \geq 3 \).
Gao et al.~\cite{gao2018exact} proposed graph reduction techniques and a BnB algorithm with a dynamic vertex selection mechanism.
Zhou et al.~\cite{zhou2021improving} offered a BnB algorithm using a graph coloring heuristic for bounding the \( k \)-plex size.
Jiang et al.~\cite{jiang2021new} designed a partition-based upper bound for BnB algorithms.
Chang et al.~\cite{chang2022efficient} introduced a matrix-based BnB algorithm incorporating both first and second-order pruning approaches.
%
Within the scope of these algorithms, aside from the work~\cite{xiao2017fast} which presents a non-trivial best time complexity of \( O(c^n n^{O(1)}) \) where \( 1.94 \leq c < 2 \) for $k\geq 3$, the complexities of other algorithms remain trivial at \( O(2^n n^{O(1)}) \). 
To the best of our knowledge, our work is the first to reduce the time complexity of    MKP   to \( O(2^{n/2} n^{O(1)}) \).

\textbf{Approximation for MKP.}
Heuristic approaches for MKP have also been developed to find near-optimal solutions. 
Gujjula and Balasundaram~\cite{gujjula2014hybrid} combined greedy randomized adaptive search (GRASP) with tabu search to tackle MKP.
Miao et al.~\cite{miao2017approaches} enhanced the GRASP method by improving the initial solution construction process.
Zhou et al.~\cite{zhou2017frequency} proposed a frequency-driven tabu search heuristic for MKP, utilizing multiple neighborhood data.
Chen et al.~\cite{chen2020local} developed a local search strategy based on double-attributes incremental neighborhood updating and dynamic configuration checking.
Pullan~\cite{pullan2021local} proposed a local search heuristic focused on avoiding search cycles in MKP.
Although these approaches are all approximation algorithms, the difference with annealing-based algorithms (both classical and quantum) is that the runtime of an annealing-based algorithm is provided as an input parameter and the solution is progressively refined. Therefore, we select the classical simulated annealing algorithm and the state-of-the-art MILP optimizer Gurobi as baselines for a direct comparison with qaMKP in terms of efficiency and effectiveness.

\textbf{Quantum Database Algorithms.}
Research on quantum database algorithms has increased significantly. \c{C}alikyilmaz et al.~\cite{10.14778/3598581.3598603} studied the potential short and long-term impacts of quantum algorithms on databases. 
Trummer and Koch~\cite{10.14778/2947618.2947621} optimized multiple queries using an adiabatic quantum annealer. Sch{\"o}nberger~\cite{schonberger2022applicability} developed quantum algorithms to determine optimal join orders in databases, which led to further work by Winker et al.~\cite{winker2023quantum} on a variational quantum circuit and by Nayak et al.~\cite{nayak2023constructing} on an algorithm using a quantum annealer. 
In graph databases, quantum computing offers new possibilities. Childs and Goldstone~\cite{childs2004spatial} investigated quantum walks for graph navigation, and Paparo and Martin-Delgado~\cite{paparo2013google} suggested that quantum PageRank algorithms may outperform classical ones. McGeoch and Wang~\cite{mcgeoch2013experimental} applied quantum annealers to graph-related problems. Schuld and Killoran~\cite{schuld2021quantum} and Bai et al.~\cite{bai2021learning} proposed quantum machine learning methods for processing graph data. 
The maximum clique problem has been addressed with Grover's search by Boji{\'c}~\cite{bojic2012quantum}, and Chang et al.~\cite{chang2018quantum} designed specific quantum circuits for such searches. Other models, such as quantum adiabatic evolution by Childs et al.~\cite{childs2002finding} and annealing by Chapuis et al.~\cite{chapuis2019finding}, were explored for clique problems, though their applications are limited across different problem types.
For the \(k\)-clique problem, Childs and Eisenberg~\cite{childs2005quantum} and Metwalli et al.~\cite{metwalli2020finding} used quantum subset search and Grover's algorithm for small \(k\). 
These techniques are not suitable for the MKP because relaxing clique constraints to form a \(k\)-plex makes quantum circuit representation difficult. To our knowledge, our work is the first to propose quantum algorithms for the MKP.

\section{Conclusion and Limitation}\label{sec:6}
\textbf{Conclusion.}
In this paper, we demonstrated that the Maximum $k$-Plex Problem  can be solved in \( O^*(2^{n/2}) \), achieving a near-quadratic speedup over the state-of-the-art. 
We proposed the first two gate-based quantum  algorithms, qTKP and qMKP, to attain this complexity. 
qTKP finds a \( k \)-plex of a given size by  translating the problem into quantum logic using quantum circuits and identifying the solution in the search space via an   oracle. 
qMKP utilizes binary search in conjunction with qTKP to progressively find the maximum solution. 
To exhibits higher qubit resource utilization, we further proposed the first annealing-based  algorithm  qaMKP by transforming the problem into a QUBO form with only \(O(n\log n)\) slack variables. 
We performed proof-of-principle experiments using state-of-the-art IBM gate-based quantum simulators and the D-Wave adiabatic quantum annealer to demonstrate the practicality and efficacy of our methods. 

\textbf{Limitation.}
Presently, quantum hardware has not achieved the fault-tolerant  computing level required for large-scale qubit operation~\cite{preskill2018quantum}. 
Therefore, this paper does not aim to show practical speedups for industrially relevant scenarios, as, to the best of our knowledge, no prior research using QPUs for databases or any other purpose has achieved this.
We believe it may not be ideal to wait for fully operational hardwares before designing suitable quantum algorithms for  database problems. Studying the asymptotic behavior of quantum   algorithms on small datasets under current hardware constraints remains valuable for understanding how database problems may evolve under new computational paradigms. Therefore, this work  provides solid theoretical support for  data  problems at this early stage of research and we hope this work will attract more interest from the   community.

\section*{Acknowledgments}
This work is supported in part by a Singapore MOE AcRF Tier-2 grant MOE-T2EP20221-0015 and a Singapore MOE AcRF Tier-1 project RT6/23. 

\bibliographystyle{IEEEtran} 
\bibliography{IEEEabrv,IEEEexample}


%


\begin{IEEEbiography}
[{\includegraphics[width=1in,height=1.25in,clip,keepaspectratio]{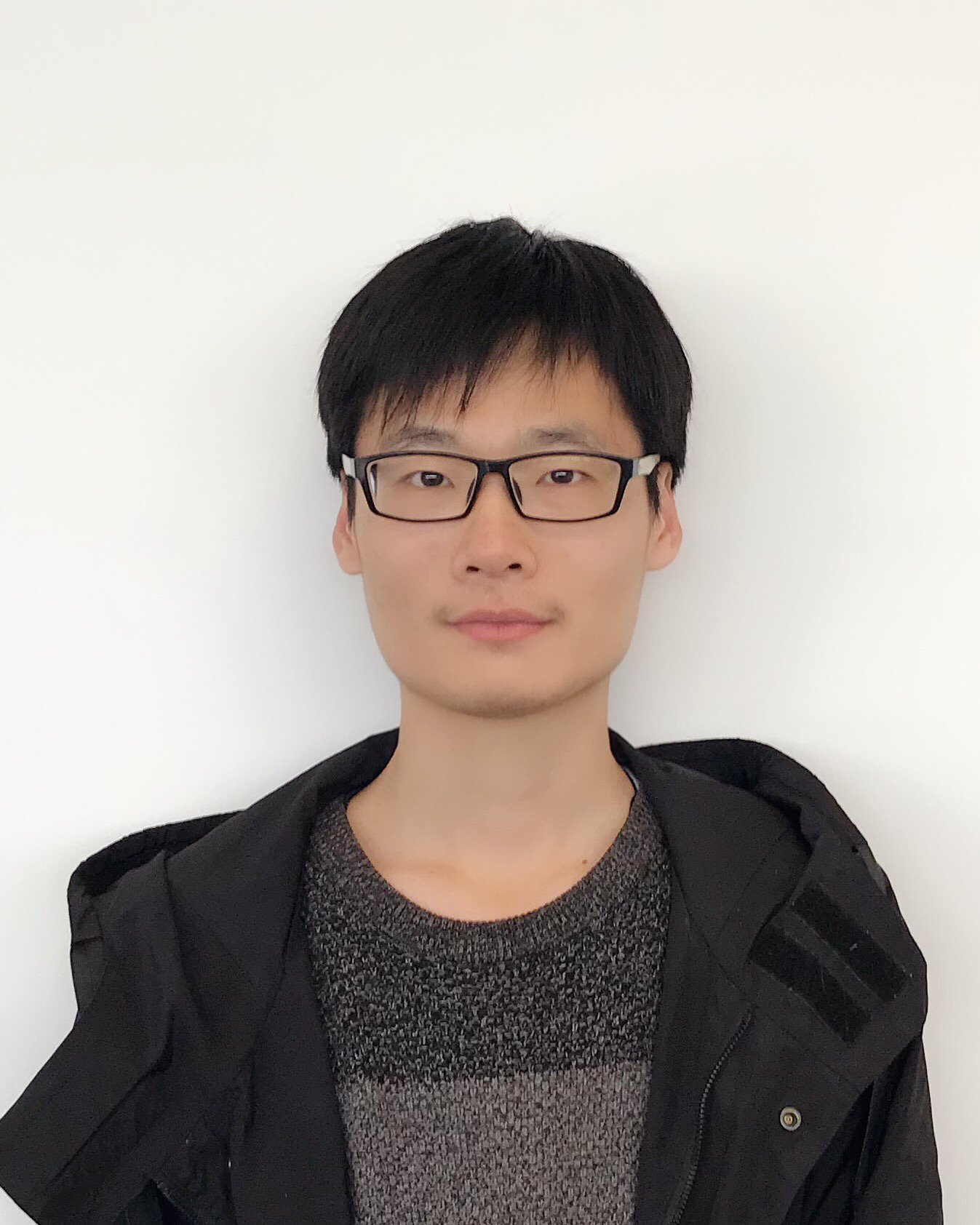}}]
{Xiaofan Li}
 received his BS degree from Wuhan University, China, in 2016, and his PhD degree from Swinburne University, Australia, in 2022. 
 He is currently a postdoctoral researcher at Nanyang Technological University, Singapore. 
 His research interests include graph database, graph neural network, and quantum database algorithm.
\end{IEEEbiography}
\vspace{-20pt}

\begin{IEEEbiography}
[{\includegraphics[width=1in,height=1.25in,clip,keepaspectratio]{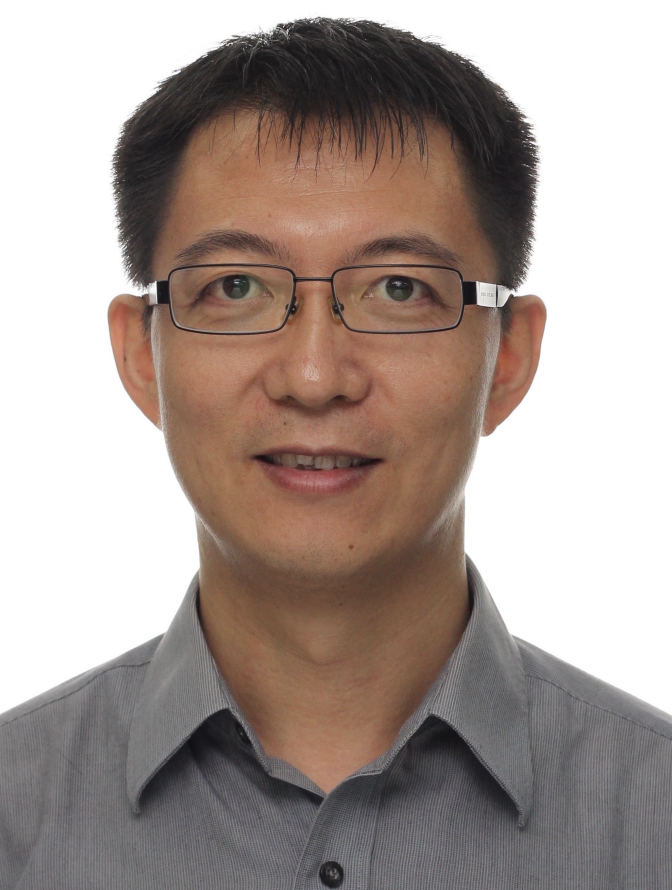}}]
{Gao Cong}
 is currently a Professor in the College of Computing and Data Science at Nanyang
Technological University (NTU). He previously
worked at Aalborg University, Denmark, Microsoft
Research Asia, and the University of Edinburgh.
He received his PhD degree from the National University of Singapore in 2004. His current research
interests include spatial data management, ML4DB,
spatial-temporal data mining, and recommendation
systems. 
\end{IEEEbiography}
\vspace{-20pt}

\begin{IEEEbiography}[{\includegraphics[width=1in,height=1.25in,clip,keepaspectratio]{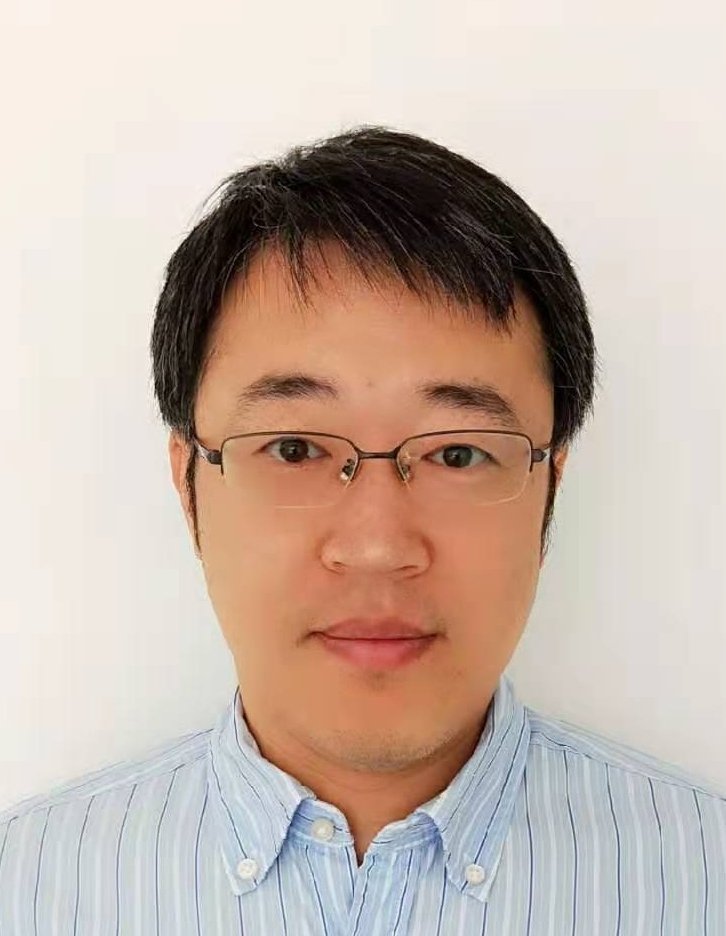}}]{Rui Zhou}
received the BS and MS degrees
from Northeastern University, China, in 2004 and
2006, respectively, and the PhD degree from the
Swinburne University of Technology, Australia, in
2010. He is currently a senior lecturer with the Swinburne
University of Technology, Australia. His research
interests include database, data
mining and algorithms. 
\end{IEEEbiography}


\vspace{11pt}


\vfill

\end{document}